\documentclass[aps,prb,twocolumn,superscriptaddress]{revtex4-1}
\usepackage{amsmath}
\usepackage{mathtools}
\usepackage{xcolor}
\usepackage{graphicx}

\usepackage{simplewick}
\usepackage{bbold}
\usepackage[title]{appendix}
\usepackage{natbib}
\usepackage{hyperref}
\usepackage{chngcntr}

\newcommand{\vecbf}[1]{{\bf #1}}

\begin{document}

\title{Numerically exact mimicking of quantum gas microscopy for interacting lattice fermions}
\author{Stephan Humeniuk}
\email{stephan.humeniuk@gmail.com}
\affiliation{Institute of Physics, Chinese Academy of Sciences, Beijing 100190, China}
\author{Yuan Wan}
\affiliation{Institute of Physics, Chinese Academy of Sciences, Beijing 100190, China}
\affiliation{University of Chinese Academy of Sciences, Beijing 100049, China}
\affiliation{Songshan Lake Materials Laboratory, Dongguan, Guangdong 523808, China}

\begin{abstract}

A numerical method is presented for reproducing fermionic quantum gas microscope experiments in equilibrium. 
By employing nested componentwise direct sampling of fermion pseudo-density matrices, 
as they arise naturally in determinantal quantum Monte Carlo (QMC) simulations, a stream 
of pseudo-snapshots of occupation numbers on large systems can be produced.  
There is a sign problem even when the conventional determinantal QMC algorithm can be made sign-problem free,
and every pseudo-snapshot comes with a sign and a reweighting factor. 
Nonetheless, this ``sampling sign problem'' turns out to be weak and manageable in a large, relevant parameter regime.
The method allows to compute distribution functions of arbitrary quantities defined in occupation number space
and, from a practical point of view,
facilitates the computation of complicated conditional correlation functions. 
While the projective measurements in quantum gas microscope experiments achieve
\emph{direct} sampling of occupation number states from the density
matrix, the presented numerical method requires a Markov chain as an
intermediate
step and thus achieves only \emph{indirect} sampling,
but the full distribution of pseudo-snapshots after (signed)
reweighting is identical to the distribution
of  snapshots from projective measurements

\end{abstract}
\maketitle
  
\section{Introduction}

The Hubbard model is a highly simplified, yet paradigmatic model of materials with strong correlations 
which has found an accurate physical realization in cold atomic gases in optical lattices \cite{Gross2017}. 
Its phase diagram is still poorly understood, which has led to an intense 
synergy of numerical approaches \cite{LeBlanc2015HubbardBenckmark, Schaefer2020MultiMethod}.

Remarkably, fermionic quantum gas microscopes \cite{Cheuk2015, Haller2015single, Parsons2015Site-Resolved, Omran2015Microscopic, Edge2015Imaging, Brown2017Spin} 
[see also Refs.~\cite{Hartke2020, Koepsell2020} and references therein]
with single-site and single-atom resolution
give access to the full distribution function of occupation number states. 
This has allowed the direct measurement of two-point correlation functions \cite{Cheuk2016Observation, Boll2016, Parsons2016} and of
more unconventional quantities 
such as the full counting statistics (FCS) of macroscopic operators \cite{Mazurenko2017} 
or the non-local string order parameter characterizing
spin-charge separation \cite{Hilker2017, Salomon2018, Vijayan2020} in 1D.
Conditional correlation functions around dopants \cite{Koepsell2019, Koepsell2020Evolution}
and the analysis of patterns in the snapshots \cite{Chiu2019} have revealed polarons in the doped Hubbard model, 
in and out of equilibrium \cite{Ji2020DynamicalInterplay}. 
Furthermore, time-dependent measurements give access to transport properties \cite{Nichols2019Spin, Brown2019BadMetallic, Anderson2019Conductivity}.
In this context, comparison with numerical simulations 
is not only important for calibrating e.g. the temperature 
in cold atoms experiments, but quite generally for reliable benchmarking
to prepare quantum simulators for parameter regimes where classical simulations are impossible. 

Yet, for fermions in $d \ge 2$ dimensions, a numerically exact technique 
for mimicking such projective measurements of occupation number shapshots is still missing.
A single hole in a system of infinitely strongly repulsive fermions ($t-J$ model) can be simulated 
with a world-line loop algorithm \cite{Brunner1998, Brunner2000} without a sign problem and, more recently, 
worm algorithm Monte Carlo \cite{Prokofev1998} applied to the $t-J$ model has given unbiased 
results for spin configurations around a small number of dopants \cite{Blomquist2019AbInitio, Blomquist2020Evidence}.
However, by definition, the $t-J$ model neglects doublon-hole fluctuations,
and for the Fermi-Hubbard model at finite interaction 
path integral Monte Carlo simulations \cite{Hirsch1982, Hirsch1986} are only possible in one dimension
due to the fermionic sign problem which is extensive in the system size.
We extend the determinantal QMC (DQMC) algorithm \cite{Blankenbecler1981, Loh1992, Assaad2002}
by an inner loop, where Fock configurations are sampled \emph{directly}, i.e. without autocorrelation time,
from a fully tractable quasiprobability distribution. 
A common technique for obtaining a tractable joint probability distribution, which can be sampled directly,
is to model it as a product of conditional distributions \cite{Larochelle2011NADE}, i.e.~as a directed graphical model \cite{Pearl2014probabilistic}. 
This idea is at the heart of autoregressive neural networks \cite{Larochelle2011NADE, Uria2016NADE},
the generation of natural images pixel by pixel \cite{vandenOord2016PixelRecurrent} and recent algorithms
for simulation and generative modelling of 
quantum systems \cite{Sharir2020Autoregressive, Wu2019Autoregressive, Wang2020NAQS,Ferris2012PerfectSampling, Han2018GenerativeModellingMPS, Clifford2018classical, XiaopengLi2019}.
\begin{figure}[t!]
 \includegraphics[width=0.9\linewidth]{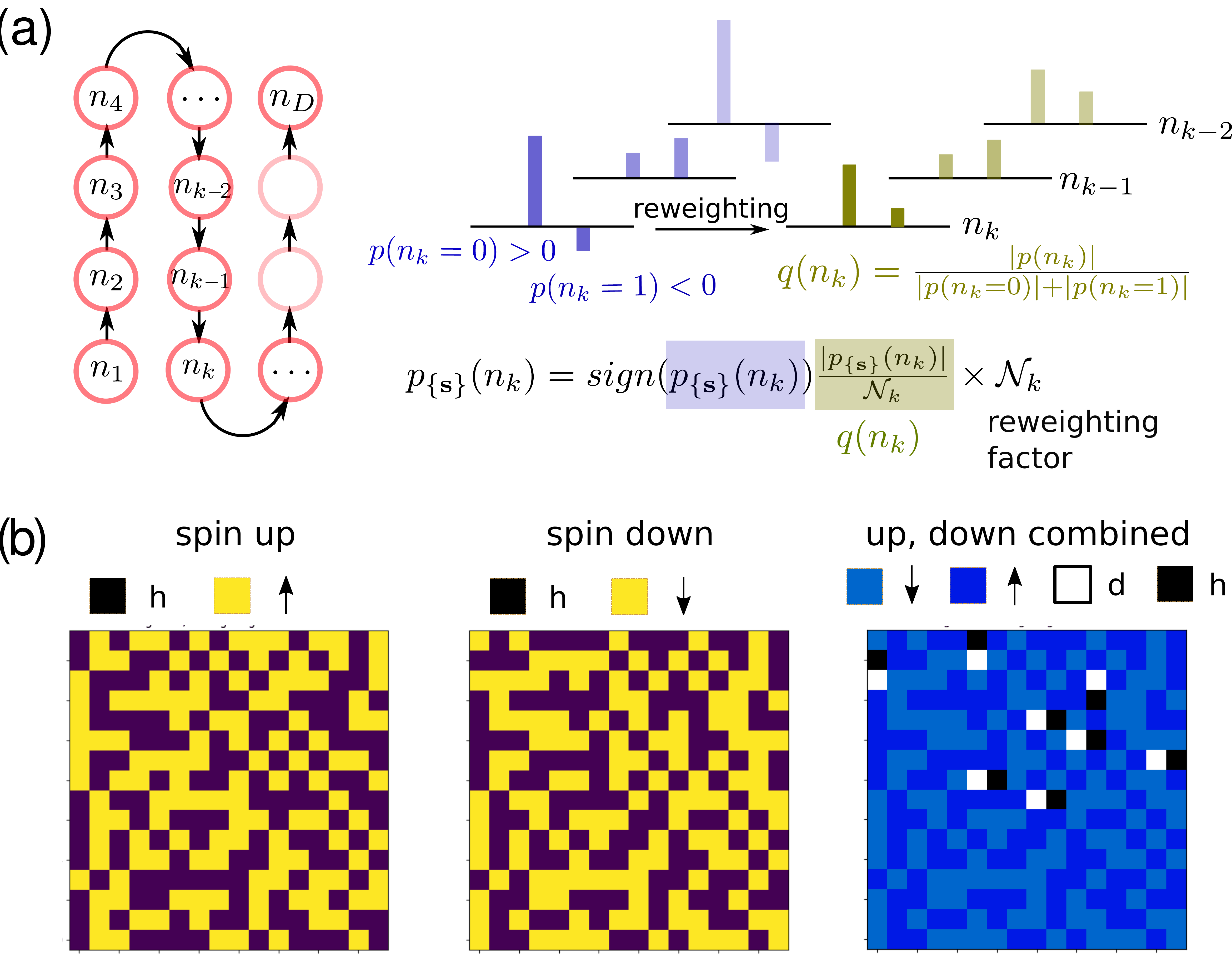}
 \caption{Componentwise direct sampling in a single configuration of Hubbard-Stratonovich fields $\{ \vecbf{s} \}$. 
 (a) Lattice sites (red circles) are ordered boustrophedonically and 
 the joint distribution of their occupation numbers is written as a chain of conditional probabilities.
 For each sampled component (i.e. lattice site $k$), the pseudo-probability distribution $p_{\{\vecbf{s}\}}(n_k)$ is reweighted
 such that samples can be drawn from a valid probability distribution $q(n_k)$.
(b) Pseudo-snapshot generated for a given Hubbard-Stratonovich field configuration. 
Doublon-hole (d-h) fluctuations, first revealed experimentally by bunching of ``anti-moments'' \cite{Cheuk2016Observation},
are clearly visible in this pseudo-snapshot, which
comes with a positive sign and a modest reweighting factor of $R \approx 1.22$.
Square lattice with parameters $U/t=10, \beta t = 10, \langle n \rangle = 1, L=16$.}
 \label{fig:sample_snapshot}
\end{figure}

Alternative DQMC approaches, summing all Fock states implicitly, exist for 
computing the FCS of quadratic operators \cite{Humeniuk2017} and all elements of the 
reduced density matrix on small probe areas \cite{Humeniuk2019}. 
The nested componentwise direct sampling technique presented here is more versatile in that pseudo-snapshots 
can be produced for probe areas as large as in current experiments with 
the proviso that a (mild) sign problem is manageable in experimentally relevant regimes. 
The resulting distribution of (a sufficiently large number of) pseudo-snapshots after 
reweighting is - within controllable statistical error - 
identical to the distribution of snapshots from projective measurements as generated in quantum gas 
microscope experiments.

\begin{figure*}
\includegraphics[width=0.95\linewidth]{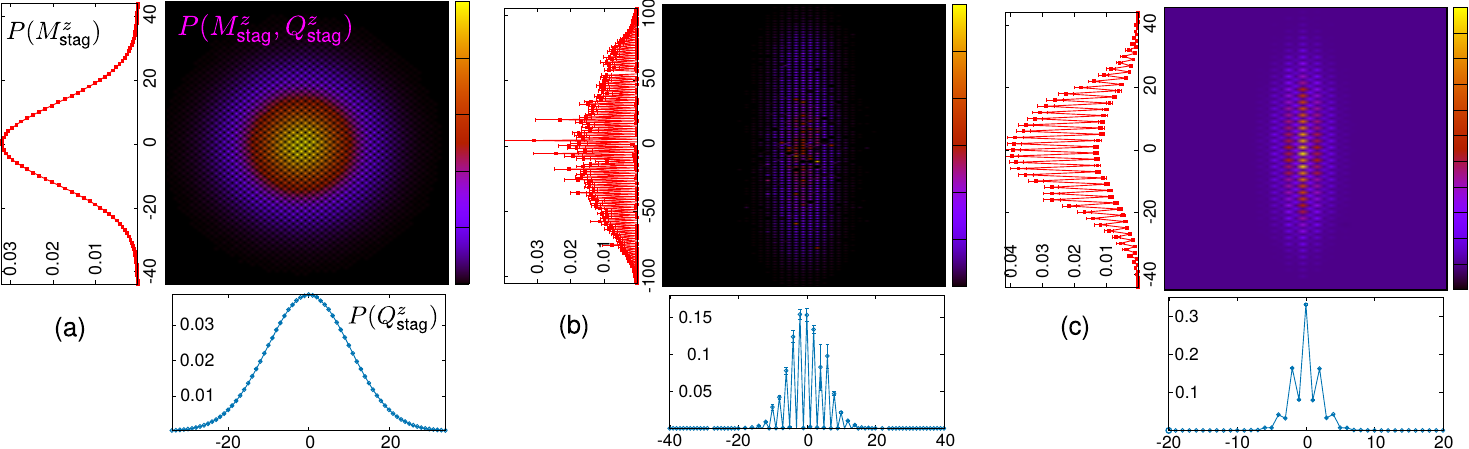}
\caption{Joint distribution $P(M^{z}_{\text{stag}}, Q_{\text{stag}}^{z})$ at half filling.
(a) Weak Hubbard interaction $U/t=1, \beta t=4, L=12, L_A=12$.
 Even-odd oscillations, which are visible in the joint distribution, are smeared out in the marginal 
 distributions (left and bottom of each panel).
(b) Close to the metal-insulator crossover: $U/t=8, \beta t = 5, L=12, L_A=12$. 
 In the heat map outliers have been set to zero.
(c) Strong Hubbard interaction: $U/t=14, \beta t=5, L=16, L_A=8$.}
\label{fig:joint_FCS}
\end{figure*}

We calculate (i) the joint FCS of the staggered spin and pseudo-spin magnetization 
of the Hubbard model at half filling, (ii) the distribution of the total number of holes and doubly occupied sites 
as a function of doping at high temperature, and (iii) the magnetization environment of a polaron, where we find qualitative agreement 
with a recent quantum Monte Carlo simulation for a single hole in the $t-J$ model \cite{Blomquist2019AbInitio}.
An apparent discrepancy between Ref.~\cite{Blomquist2019AbInitio} and the quantum gas microscope experiment of Ref.~\cite{Koepsell2019},
which we can also reproduce qualitatively, can be pinpointed to a difference in doping regimes. 

\section{Nested and componetwise direct sampling}

We are considering the single-band Hubbard model
\begin{equation}
 H = -t \sum_{\langle i,j \rangle, \sigma = \uparrow, \downarrow} (\hat{c}^{\dagger}_{i, \sigma} \hat{c}_{j, \sigma} + h.c.)
  + U \sum_{i} \hat{n}_{i,\uparrow} \hat{n}_{i,\downarrow} - \mu \sum_{i, \sigma} \hat{n}_{i, \sigma},
\end{equation}
with the usual notation, and treat it within the DQMC framework \cite{Blankenbecler1981, Loh1992, Assaad2002}:
After a Trotter-Suzuki decomposition of the density operator $\hat{\rho} \sim \exp(-\beta H)$ at inverse temperature $\beta=1/T$
into $N_{\tau} = \beta / \Delta \tau$ imaginary times slices and a Hubbard-Stratonovich (HS) transformation for decoupling 
the interactions by introducing a functional integral over HS fields $\{ \vecbf{s} \}$, the density operator reads \cite{Grover2013}
\begin{subequations}
\begin{align}
 \hat{\rho} &= \frac{1}{Z} \sum_{\{ \vecbf{s} \}} \prod_{\sigma = \uparrow, \downarrow} \left( w^{\sigma}_{\{ \vecbf{s} \}} 
 \frac{e^{-\sum_{i,j} X_{i,j}^{\sigma}(\{ \vecbf{s} \}) \hat{c}_{i,\sigma}^{\dagger} \hat{c}_{j, \sigma}}}{w^{\sigma}_{\{ \vecbf{s} \}}} \right) \\
 &\equiv \frac{1}{Z}  \sum_{\{ \vecbf{s} \}} \prod_{\sigma = \uparrow, \downarrow} w_{\{ \vecbf{s}\}}^{\sigma} \hat{\rho}_{\{ \vecbf{s}\}}^{\sigma}.
\end{align}
\label{eq:total_rho}
\end{subequations}
Here, formally we have $\exp(-X^{\sigma}) \equiv \prod_{l=1}^{N_{\tau}} B_{l}^{\sigma}$ where $B_{l}^{\sigma} = e^{-\Delta \tau V_l^{\sigma}(\{\vecbf{s}_l\})} e^{- \Delta \tau K}$
is the matrix representation of the single-particle propagators for spin $\sigma$ of the potential and kinetic part after HS transformation \cite{Assaad2002}. 
Note that we use the conventional HS transformation \cite{Hirsch1983} in which the discrete auxiliary fields couple to the $S^{z}$-component of the 
electron spin, $S_{i}^{z} = \hat{n}_{i,\uparrow} - \hat{n}_{i,\downarrow}$, which, as will be discussed below, is crucial.
After integrating out the fermionic degrees of freedom, $w_{\{ \vecbf{s}\}}^{\sigma} = \det\left( \mathbb{1} + e^{-X^{\sigma}(\{ \vecbf{s} \})}\right) \equiv Z(\{\vecbf{s}\})$ is the contribution 
of the spin component $\sigma$ to the Monte Carlo weight of the HS field configuration $\{ \vecbf{s} \}$,
which can also be interpreted as the partition sum of the non-interacting fermion system 
$\hat{\rho}_{\{ \vecbf{s}\}}^{\sigma}$.
Since the kinetic and potential matrices in the matrix product leading to Eq.~\eqref{eq:total_rho} do not commute, 
the resulting matrix $\frac{e^{-X^{\sigma}(\{ \vecbf{s}\})}}{Z(\{\vecbf{s}\})}$
is not Hermitian and (except in 1D) not all diagonal matrix elements of $\hat{\rho}_{\{ \vecbf{s} \}}$,
are semi-positive-definite; hence, $\hat{\rho}_{\{ \vecbf{s} \}}$ is termed a \emph{pseudo}-density matrix, while the total $\hat{\rho}$ in Eq.~\eqref{eq:total_rho}
is a true density matrix.

The structure of the density matrix in Eq.~\eqref{eq:total_rho} suggests a nested sampling approach, 
in which the HS fields of the pseudo-density matrices $\hat{\rho}_{\{ \vecbf{s} \}}$
are sampled using the Markov chain of the conventional determinantal QMC algorithm, 
while the occupation numbers can be sampled \emph{directly}, i.e. without autocorrelation time, 
from each free-fermion pseudo-density matrix $\hat{\rho}_{\{ \vecbf{s} \}}$
given that for fixed $\{ \vecbf{s} \}$ their distribution function and all its marginals can be calculated efficiently. 

From the chain rule of basic probability theory every probability distribution 
can be decomposed into a chain of conditional probabilities 
\begin{equation}
 p_{\{ \vecbf{s} \}}(n_1, n_2, \ldots, n_D) = \prod_{k=1}^{D}p_{\{ \vecbf{s} \}}(n_k | n_{k-1}, n_{k-2}, \ldots, n_1)
 \label{eq:ancestral_sampling1}
\end{equation}
where some ordering of the random variables $n_1, n_2, \ldots, n_D$ is implied. 
A sample from the joint distibution is then generated by traversing the chain as 
 $p_{\{ \vecbf{s} \}}(n_1) \xrightarrow{n_1 \sim p_{\{ \vecbf{s} \}}(n_1)} p_{\{ \vecbf{s} \}}(n_2 | n_1) \xrightarrow{n_2 \sim p_{\{ \vecbf{s} \}}(n_2)} p_{\{ \vecbf{s} \}}(n_3| n_2, n_1) \rightarrow \ldots$, 
where $n_k \sim p_{\{ \vecbf{s}\}}(n_k)$ denotes sampling variable $n_k$ from $p_{\{ \vecbf{s} \}}(n_k|n_{k-1}, n_{k-2}, \ldots)$
and the sampled value is ''inserted`` into the next conditional probability along the chain [see Fig.~\ref{fig:sample_snapshot}(a)].

Below we discuss how to calculate the conditional 
quasiprobabilities in Eq.~\eqref{eq:ancestral_sampling1} 
for a free fermion pseudo-density matrix $\rho_{\{ \vecbf{s} \}}^{\sigma}$. 
Note that for fixed HS field configuration, the pseudo-density matrices for spin up and 
down are statistically independent. Per HS sample, spin up and down 
occupancies are sampled independently and then 
combined into full pseudo-snapshots, see Fig.~\ref{fig:sample_snapshot}(b).
Henceforth, we drop the subscripts $\{\vecbf{s}\}$ and $\sigma$ 
for notational convenience.

\subsection{Direct sampling in the grand canonical ensemble}

In the atomic microscopy, the quantity of central interest is the quasiprobability to 
find a given snapshot of the fermion occupation:
\begin{align}
p(n_1,n_2,\cdots n_D) = \mathrm{Tr}(\rho \Pi_{n_1}\Pi_{n_2}\cdots \Pi_{n_D}),
\label{eq:joint_quasiprob}
\end{align}
where $n_i = 0,1$ is the occupation number on a given site $i$. The projectors $\Pi_{n_i = 0} = \hat{c}_i \hat{c}^\dagger_i$ 
and $\Pi_{n_i = 1} = \hat{c}^\dagger_i \hat{c}_i$ project onto the Fock states with occupation number $n_i$. 
We may think of $n_1,n_2\cdots n_D$ as an ensemble of $D$ random binary variables. 
Provided that we can easily compute the marginals $p(n_1) = \mathrm{Tr}(\rho \Pi_{n_1})$, $p(n_1,n_2) = \mathrm{Tr}(\rho \Pi_{n_1}\Pi_{n_2})$,
etc. and thus the conditional quasiprobabilities,
we may sample $n_1,n_2\cdots n_D$ by a componentwise direct sampling. 

In the grand canonical ensemble all marginal quasiprobability distributions of the occupation 
numbers can be computed straightforwardly, namely
\begin{widetext}
\begin{align}
p(n_1,n_2,\cdots n_k) = (-)^{n_1+n_2+\cdots n_k}
\det\begin{pmatrix}
G_{11}-n_1 & G_{12} & \cdots & G_{1k}\\
G_{21} & G_{22}-n_2 & \cdots & G_{2k}\\
\vdots & \vdots & \ddots & \vdots \\
G_{k1} & G_{k2} & \cdots & G_{kk}-n_k
\end{pmatrix},
\label{eq:joint_quasiprob_marginal}
\end{align}
\end{widetext}
where $G_{ij} \equiv G_{ij}^{\sigma}({\{\vecbf{s}\}}) = \langle c_{i,\sigma} c_{j,\sigma}^{\dagger}\rangle_{\{ \vecbf{s} \}}$ is the equal-time single-paricle Green's function 
of spin species $\sigma$ for a given HS field configuration $\{ \vecbf{s} \}$ at a randomly chosen imaginary time slice. 
Eq.~\eqref{eq:joint_quasiprob_marginal} is proven in appendix \ref{app:proof}.

We now decompose the high-dimensional quasiprobability distribution Eq.~\eqref{eq:joint_quasiprob} into a chain of conditional 
quasiprobability distributions, which by definition can be computed as 
\begin{align}
p(n_{k+1} | n_1,n_2\cdots n_k) = \frac{p(n_1,n_2\cdots n_k,n_{k+1})}{p(n_1,n_2,\cdots n_k)}.
\end{align}
Inserting Eq.~\eqref{eq:joint_quasiprob_marginal} and using the determinant formula for block matrices we find:
\begin{subequations}
\begin{align}
p(0|n_1,n_2\cdots n_k) = G_{k+1,k+1}-\Delta G_{k+1}, \\
p(1|n_1,n_2\cdots n_k) = 1-G_{k+1,k+1}+\Delta G_{k+1},
\end{align}
\label{eq:component_quasiprob}
\end{subequations}
with the ``correction term"
\begin{align}
\Delta G_{k+1} = \sum^k_{i=1} G_{k+1,i} (G_{K,K}-N_{K,K})^{-1}G_{i,k+1},
\label{eq:correction_term}
\end{align}
and where $K=\{1,2,...,k\}$ denotes the ordered set of site indices,
$G_{K,K}$ is the corresponding submatrix of the Green's function, and 
$N_{K,K}=\text{diag}(n_1, n_2, ..., n_k)$ is a diagonal matrix whose entries 
are the sampled occupation numbers on the sites $K$.

If the correction term $\Delta G_{k+1}$ were zero, then the conditional probability
$p(n_{k+1} | n_1, n_2, \ldots)$ would be simply given by the diagonal element $G_{k+1, k+1}$ of the Green's function 
and be independent of the other occupation numbers ~\footnote{An approximation which 
considers only the diagonal elements \cite{Khatami2020Visualizing} of the Green's function 
may give a particle number distribution with the correct average and variance, but 
all correlations between sites in a given HS sample will be lost.}.
Therefore the correction term is crucial for inter-site correlations.
While traversing the chain of conditional probabilties, the block structure of the matrix whose inverse is required in Eq.~\eqref{eq:correction_term} can be exploited recursively
such that no calculation of a determinant or matrix inverse from scratch is necessary (see appendix \ref{app:block_determinant}).

\begin{figure*}[t!]
\includegraphics[width=1.0\textwidth]{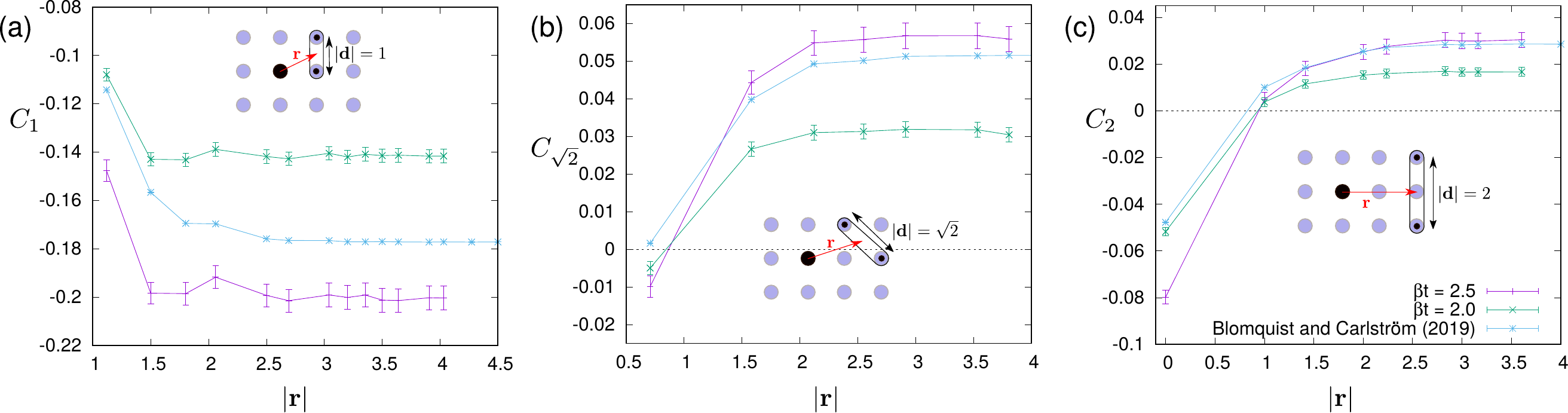}
\caption{Three-point spin-charge correlations $C_{|\vecbf{d}|}(|\vecbf{r}|)$, as depicted in the insets,  around 
an \emph{isolated} hole (see main text). $U/t=14$, $\mu/t = -3$, $\beta t \in \{2, 2.5\}$, system size
$L \times L$ with $L=10$. A total number of $10^8$ pseudo-snapshots has been generated with 224 independent Markov chains.
For $\mu/t = -3$ and $L=10$, the average number of excess holes is $\langle N_h \rangle - \langle N_d \rangle \approx 0.35$ 
(see FCS of $N_h$ and $N_d$ in appendix \ref{app:FCS_doublons_holes}).
The data is consistent with data for comparable parameters of temperature
and interactions ($\beta t = 2.2 $ and $\beta J = 0.66 $) from Ref.~\cite{Blomquist2019AbInitio}
where a single hole in the $t-J$ model was simulated.}
\label{fig:three_point_spincharge_corr}
\end{figure*}

\subsection{Reweighting}
\label{sec:reweighting}
As said earlier, the pseudo-density operator $\hat{\rho}_{\{\vecbf{s}\}}$ is not Hermitian and 
not all conditional quasiprobabilities in Eq.~\eqref{eq:component_quasiprob} are non-negative.
Therefore, we rewrite them as
\begin{equation}
 p(n_k) = \text{sign}(p(n_k)) \frac{|p(n_k)|}{\mathcal{N}_k} \times \mathcal{N}_k.
\end{equation}
with the shorthand notation \mbox{$p(n_k) \equiv p(n_k | n_1, n_2, \ldots, n_{k-1})$}.
The sampling for component $k$ is then carried out using the 
valid probability distribution \mbox{$q(n_k) \equiv \frac{|p(n_k)|}{\mathcal{N}_k}$}
with normalization $\mathcal{N}_k = |p(n_k=0)| + |p(n_k=1)|$ [see Fig.~\ref{fig:sample_snapshot}(a)].

Having sampled the entire chain of conditional probabilities for both spin components, 
the generated snapshot is associated with a (signed) reweighting factor $R= R^{\uparrow} R^{\downarrow}$ where
\begin{equation}
 R^{\sigma} =  \text{sign}(w^{\sigma}_{\{ \vecbf{s}\}})   \prod_{k=1}^{D} \left( \text{sign}(p_{\{ \vecbf{s} \}}(n^{\sigma}_k)) \, \mathcal{N}^{\sigma}_k \right),
\end{equation}
and all quantities $\mathcal{O}(\vecbf{n}) = \mathcal{O}(n_1, n_2, \ldots, n_D)$ that are evaluated on $M$ generated snapshots $\{ \vecbf{n}_i \}_{i=1}^{M}$ 
need to be reweighted as 
\begin{equation}
 \langle \mathcal{O} \rangle = \frac{\sum_{i=1}^{M} R_i  \mathcal{O}(\vecbf{n}_i)}{\sum_{i=1}^{M} R_i}.
\end{equation}

The joint pseudo-probabilitiy $p(n_1, n_2, \ldots)$ can be factored in arbitrary order into components. However, 
the reweighted distribution $q(n_1, n_2, \ldots)$ and thus the magnitude and sign of the 
reweighting factor depends on the chosen factor ordering,
leaving room for optimization in a given HS sample.

The severity of the sign problem is basis dependent: We find that it strongly depends on the single-particle basis for sampling
and the chosen HS transformation. Sampling in the $S^{x}$- or $S^{y}$-basis of the electron spin 
(or in the $S^{z}$-basis in momentum space) leads to a very strong sign (phase) problem. 
Chosing the HS transformation that couples to the electron charge density \cite{Assaad2002} rather than the electron spin
also gives a very severe sign problem. 

Even at half filling, where the DQMC algorithm can be made sign-problem free for the purpose of computing expectation values, 
there is a sampling sign problem, with a non-uniform dependence on $U/t$.
The average sign diminishes as $U/t$ increases from $U/t=0$ 
and reaches a minimum at an intermediate value $(U/t)_{\text{MIC}} \approx 4-7$,
where a metal-to-insulator crossover \footnote{
In the Hubbard model at half filling the charge gap 
scales as $e^{-2 \pi \sqrt{t/U}}$ for $U/t\ll 1$ due to an exponentially diverging 
antiferromagnetic correlation length and becomes $U/2$ for $U/t \gg 1$.
When the temperature is larger than the charge gap, the system 
becomes metallic. Then, keeping the temperature constant, there is a 
metal-to-insulator crossover as a function of U/t at some $(U/t)_{\text{MIC}}$.
It is to be expected from the foregoing argument that this $(U/t)_{\text{MIC}}$ increases with 
increasing temperature, which is consistent with the shift 
of the minimum of $< \tilde{s} >$ as a function of temperature 
in Fig.~\ref{fig:av_sign_halffilling} in appendix \ref{app:sign_problem}.
} occurs \cite{Kim2020HalffilledHubbard}. 
For $U/t > (U/t)_{\text{MIC}}$ the sign problem again gradually 
becomes much less severe. The dependence of the sampling sign problem on temperature, interaction strength,
and doping is presented in Figs. \ref{fig:av_sign_halffilling}, \ref{fig:av_sign_doped_U}, and \ref{fig:av_sign_doped_temp} in Appendix C.
The average sampling sign of pseudo-snapshots for the parameters in all figures of 
the main text is $\langle \tilde{s} \rangle \ge 0.75$ with $\langle \tilde{s} \rangle$ defined in Eq.~\eqref{eq:average_sampling_sign}.

In all simulations, we use a Trotter discretization of $\Delta \tau t = 0.02$ and generate around 20 
pseudo-shapshots per HS sample on equidistant imaginary time slices. 
The simulation code has been verified by comparing with 
exact diagonalization results (see appendix \ref{app:benchmark_pentagon}).

\section{Applications}

\subsection{Joint full counting statistics (FCS)}

At half filling, the Hubbard model has an enlarged $(SU(2) \times SU(2)) / Z_2 = SO(4)$ symmetry \cite{CNYang1990}, 
which is the combination of spin-rotational and particle-hole symmetry
and is generated by two commuting sets of angular momentum operators describing the total spin and total pseudo-spin of the system, respectively. 
As the temperature is lowered, 
domains form with an order parameter of the same symmetry,
which, apart from fluctuating in length, can rotate \cite{Mazurenko2017, Humeniuk2017} on an $SO(4)$ sphere between antiferromagnetic, 
$s$-wave pairing and charge-density wave correlations, as one goes from one domain to a neighbouring domain. 
A joint histogram of the operators representing different components of the order parameter should reflect 
that they are projections of the same vector along different directions in order parameter space. 
Fig.~\ref{fig:joint_FCS} shows the joint distribution of the projections of the staggered magnetization 
\mbox{$M_{\text{stag}}^{z} = \sum_{i=1}^{N_A} (-1)^{\vecbf{i}} (\hat{n}_{i,\uparrow} - \hat{n}_{i,\downarrow})$}
and the staggered pseudo-spin
\mbox{$Q_{\text{stag}}^{z} = \sum_{i=1}^{N_A} (-1)^{\vecbf{i}} (\hat{n}_{i,\uparrow} + \hat{n}_{i,\downarrow} - 1)$}
on a square probe area of size $N_A = L_A^2$, as obtained from re-weighted pseudo-snapshots.
As $U/t$ increases [Fig.~\ref{fig:joint_FCS} (a-c)], the suppression of charge fluctuations manifests 
itself in the narrowing of the pseudo-spin distribution.
The even-odd effect visible in the joint distributions is due to the fact that an even number of sites can only accomodate 
an even magnetization of spin-$\frac{1}{2}$ (pseudo-spin) moments. 

Note that $P(M_{\text{stag}}^{z}, Q_{\text{stag}}^{z})$ could also be obtained using the generating function 
approach of Ref.~\cite{Humeniuk2017}. This is not true for the FCS of the number of doublons $N_{d} = \sum_{i=1}^{N_{A}} \hat{n}_{i,\uparrow} \hat{n}_{i,\downarrow}$ 
and the total number of holes $N_h = \sum_{i=1}^{N_{A}} \left(1 - \hat{n}_{i,\uparrow} \right) \left(1 - \hat{n}_{i,\downarrow} \right)$
since these operators are non-quadratic in fermionic operators. 
We find that e.g.~for $U/t=12$ and $\beta t = 4$, the FCS of $N_h$ and $N_d$ as a function of doping are accurately 
modelled by (shifted) binomial distributions (see appendix \ref{app:FCS_doublons_holes}).

\subsection{Three-point spin-charge correlator}

A three-point spin-charge correlator in the reference frame of the hole \cite{Koepsell2019} was calculated
for a single hole in the $t-J$ model at zero temperature with DMRG and trial wave functions \cite{Grusdt2019} 
and at finite (high and low) temperature using worm algorithm QMC \cite{Blomquist2019AbInitio}.
Presumably the same correlator was measured experimentally in Ref.~\cite{Koepsell2019} for the Hubbard model at large interactions. 
For the purpose of meaningful comparison between the Hubbard and $t-J$ model we calculate the slightly modified correlator:
\begin{equation}
 C_{|\vecbf{d}|}(\vecbf{r}) = \frac{\langle \mathcal{P}_{\vecbf{r}_0}^{h} S^{z}_{\vecbf{r}_0 + \vecbf{r} + \frac{\vecbf{d}}{2}} S^{z}_{\vecbf{r}_0 + \vecbf{r} - \frac{\vecbf{d}}{2}} \rangle}
    {\langle \mathcal{P}_{\vecbf{r}_0}^{h} \rangle}
\label{eq:third_order_corr}    
\end{equation}
where 
$S^{z}_{\vecbf{r}} = \hat{n}_{\vecbf{r,\uparrow}} - \hat{n}_{\vecbf{r},\downarrow}$,
and the projector 
\begin{align}
 \mathcal{P}_{\vecbf{r}_0}^{h} =
  & \hat{n}^{h}_{\vecbf{r}_0} \hat{n}^{s}_{\vecbf{r}_0 + \hat{e}_x} \hat{n}^{s}_{\vecbf{r}_0 - \hat{e}_x} \hat{n}^{s}_{\vecbf{r}_0 + \hat{e}_y} \hat{n}^{s}_{\vecbf{r}_0 - \hat{e}_y} \nonumber \\
  & \times \hat{n}^{s}_{\vecbf{r}_0 + \hat{e}_x + \hat{e}_y} \hat{n}^{s}_{\vecbf{r}_0 + \hat{e}_x - \hat{e}_y} \hat{n}^{s}_{\vecbf{r}_0 - \hat{e}_x - \hat{e}_y} \hat{n}^{s}_{\vecbf{r}_0 - \hat{e}_x + \hat{e}_y}
\label{eq:spinonly_env_projector}
\end{align}
with
$\hat{n}_{\vecbf{r}}^{h} = (1 - \hat{n}_{\vecbf{r},\uparrow})(1 - \hat{n}_{\vecbf{r},\downarrow})$
and
$\hat{n}_{\vecbf{r}}^{s} = (1 - \hat{n}_{\vecbf{r},\uparrow}) \hat{n}_{\vecbf{r}, \downarrow} + (1 - \hat{n}_{\vecbf{r},\downarrow}) \hat{n}_{\vecbf{r}, \uparrow}$
selects configurations with an empty site at position $\vecbf{r}_0$ surrounded from eight sides by spin-only states,
which serves to exclude nearest- and next-nearest neighbour doublon-hole pairs from the statistics. 
The conditional correlation functions in the reference frame of the hole, 
Eq.~\eqref{eq:third_order_corr},
are implemented straightforwardly by applying a filter to the pseudo-snapshots, 
whereas implementing 
such higher order correlation functions using a generalized form of Wick's theorem would require separate coding 
for each specific correlator thus hampering quick experimentation 
(although this would give better statistics since all Fock state are summed implicitly).

Fig.~\ref{fig:three_point_spincharge_corr} shows overall qualitative agreement of $C_{|\vecbf{d}|}(\vecbf{r})$
for our data for the Hubbard model and that of Ref.~\cite{Blomquist2019AbInitio} for the $t-J$ model, 
with some notable differences in $C_{\sqrt{2}}$ and $C_{2}$ 
in the immediate vicinity of the hole. A careful comparison of the $t-J$ model with the Hubbard model
would require a renormalization of all correlators in the former by a polynomial in $t/U$ \cite{Delannoy2005}
(although certain qualitative model differences may not be captured perturbatively \cite{Choy2005}).

There are qualitative differences to the experimental data 
of Ref.~\cite{Koepsell2019}, which were already noted in Ref.~\cite{Blomquist2019AbInitio}. 
The data of Ref.~\cite{Koepsell2019} can be reproduced by nested componentwise sampling, pointing,
however, to a very different conclusion: namely, that magnetic polarons may
have disappeared for the relatively high doping level of Ref.~\cite{Koepsell2019}. 
This is illustrated in appendix \ref{app:additional_data}).

\section{Conclusion}

In conclusion, the presented method for generating pseudo-snapshots 
allows both theorist to take part in the exploration of fermionic quantum microscopy 
and experimentalists to use numerical simulations in a more versatile way \cite{Code_URL}.
The data analysis of pseudo-snapshots is quite analogous to that of experimental snapshots generated 
by projective measurements except that a signed reweighting factor needs to be taken into account.
While it is not meaningful to compare individual pseudo-snapshots with actual experimental snapshots,
the full distribution of pseudo-snapshots after reweighting is identical to the distribution of snapshots from projective 
measurements. The difference is that quantum gas micropscope experiments achieve \emph{direct} sampling of occupation number states 
from the density matrix, whereas the method presented here relies on \emph{indirect} sampling from the overall
interacting fermion density matrix (we can achieve direct sampling only at the level of the constituent 
free fermion density matrices).
Arbitrary quantities can be evaluated on the reweighted pseudo-snapshots, including those that cannot be 
feasibly expressed as expectation values of operators. A case in point is the FCS of macroscopic operators which 
are higher than quadratic in fermionic operators; there, the generating function method of Ref.~\cite{Humeniuk2017}
does not apply, and evaluation as a sum of projectors onto Fock states 
would require a number of terms which is exponential in the number of sites.
 
Our nested componentwise direct sampling method is generic to all fermionic Monte Carlo methods 
that are based on the ``free fermion decomposition'' \cite{Grover2013} and 
is easily adapted to projector DQMC \cite{Assaad2002} for accessing zero-temperature properties 
where Slater determinants rather than thermal free-fermion 
pseudo-density matrices are sampled directly in the inner loop of the Markov chain. For Hubbard models
around intermediate interaction strength $U/t \approx (U/t)_{\text{MIC}}$ further improvements are required 
to reduce the sampling sign problem. There, the interest lies 
in potentially observing non-Gaussian fluctuations \cite{Moreno-Cardoner2016} 
or characterizing attraction and spin-correlations between dopants \cite{Blomquist2020Evidence} 
in the crossover from a polaronic metal to a Fermi liquid \cite{Koepsell2020Evolution}.
While a general solution of the sampling sign problem is unlikely, it remains to be investigated 
how more general HS decouplings and representations of the electron operator \cite{ZiXiangLi2019}, which were successful at eliminating 
the sign problem of the Monte Carlo weights, affect the sampling sign problem
and whether snapshots in another single-particle basis can be generated.

The total computing time spent on the generation of Figs.~1-3
amounts to the equivalent of approximately $3 \times 10^4$ CPU hours 
on an Intel(R) Core(TM) i5-6300U CPU with 2.40GHz clock cycle.
  
\section{Acknowledgments} 
We thank Xiaopeng Li and Lei Wang for motivating discussions
and acknowledge Emil Blomquist and Johan Carlstr\"{o}m for providing the raw data for Fig.~\ref{fig:three_point_spincharge_corr}
as well as comments on the manuscript. 
This work is supported by the 
International Young Scientist Fellowship from the Institute of Physics, 
Chinese Academy of Sciences, Grant No. 2018004 (Humeniuk) 
and by the National Science Foundation of China, Grant No. 11974396, and
the Strategic Priority Research Program of the Chinese Academy 
of Sciences, Grant No. XDB33020300 (Wan).
The simulations were carried out on \mbox{TianHe-1A} at the National Supercomputer Center in
Tianjin, China.

%

 
\appendix 
\counterwithin{figure}{section}
\widetext

\section{Inductive proof of Eq.~(5) in the main text} 
\label{app:proof}

Let 
\begin{equation}
G^{(0)}_{i_{\alpha}, j_{\beta}} = \langle \hat{c}_{i_{\alpha}} \hat{c}_{j_{\beta}}^{\dagger} \rangle_0 
\label{eq:def_G0}
\end{equation}
be the single-particle Green's function of a free fermion system, and $\alpha, \beta \in \{1,\ldots, D\}$. 
First we prove the well-known fact that 
Wick's theorem for higher-order correlation functions can be expressed in the compact determinant  form
\begin{equation}
 \left\langle \left(\hat{c}_{i_1} \hat{c}_{j_1}^{\dagger} \right)  \left(\hat{c}_{i_2} \hat{c}_{j_2}^{\dagger}\right) \cdots  \left(\hat{c}_{i_n} \hat{c}_{j_n}^{\dagger} \right) \right\rangle_0 
          = \det\left( G^{(0)}_{I=\{i_1, i_2, \ldots, i_n \}; J=\{j_1, j_2, \ldots, j_n \}}\right) \equiv \left[ G^{(0)} \right]_{I;J}
 \label{eq:Wick_determinant}
\end{equation}
We use the convention that $[A]_{I;J}$ refers to the determinant of the submatrix of $A$ whose row index (column index) runs in the set $I$ ($J$),
which is the ``inclusive'' definition of the minor. 

The proof goes by induction. The case $n=1$ is true by virtue of the definition \eqref{eq:def_G0}.
In the induction step, we use Wick's theorem writing all non-vanishing contractions
for a product of $n+1$ pairs of fermionic operators as
\begin{align}
&\left\langle \left( \hat{c}_{i_1} \hat{c}_{j_1}^{\dagger} \right) \cdots \left( \hat{c}_{i_{n+1}} \hat{c}_{j_{n+1}}^{\dagger} \right) \right\rangle_0
 = - G^{(0)}_{i_1, j_{n+1}} \left\langle \left(\hat{c}_{i_{n+1}} \hat{c}_{j_1}^{\dagger} \right) \left( \hat{c}_{i_2} \hat{c}_{j_2}^{\dagger} \right) \cdots \left( \hat{c}_{i_n} \hat{c}_{j_n}^{\dagger} \right) \right \rangle_0 \nonumber \\
 &- \sum_{k=2}^{n-1} G^{(0)}_{i_k, j_{n+1}} \left \langle \left( \hat{c}_{i_1} \hat{c}_{j_1}^{\dagger} \right) \left( \hat{c}_{i_2} \hat{c}_{j_2}^{\dagger} \right)  \cdots 
                   \left( \hat{c}_{i_{n+1}} \hat{c}_{j_k}^{\dagger} \right) \left( \hat{c}_{i_{k+1}} \hat{c}_{j_{k+1}}^{\dagger} \right) \cdots 
                   \left( \hat{c}_{i_{n}} \hat{c}_{j_n}^{\dagger} \right)         \right \rangle_0  \nonumber \\
 &- G^{(0)}_{i_n, j_{n+1}} \left\langle \left( \hat{c}_{i_1} \hat{c}_{j_1}^{\dagger} \right) \cdots \left( \hat{c}_{i_{n-1}} \hat{c}_{j_{n-1}}^{\dagger} \right) \left( \hat{c}_{i_{n+1}} \hat{c}_{j_{n+1}}^{\dagger} \right)\right\rangle_0 \nonumber \\
 &+ G^{(0)}_{i_{n+1}, j_{n+1}} \left \langle \left( \hat{c}_{i_1} \hat{c}_{j_1}^{\dagger} \right) \cdots \left( \hat{c}_{i_n} \hat{c}_{j_n}^{\dagger} \right) \right \rangle_0.
 \label{eq:after_generalized_Wick}
\end{align}
The minus-sign, e.g. in the second line of Eq.~\eqref{eq:after_generalized_Wick}, 
comes from the permutation of $\hat{c}_{i_{n+1}}$ with $\hat{c}_{j_k}^{\dagger}$.
Using the induction hypothesis Eq.~\eqref{eq:Wick_determinant}, 
valid for $n$ pairs of fermionic operators,
the remaining correlators in Eq.~\eqref{eq:after_generalized_Wick}
can be expressed as determinants
\begin{align}
 \left\langle \left( \hat{c}_{i_1} \hat{c}_{j_1}^{\dagger} \right) \cdots \left( \hat{c}_{i_{n+1}} \hat{c}_{j_{n+1}}^{\dagger} \right) \right\rangle_0 = 
 &- G_{i_1, j_{n+1}}^{(0)} \cdot\det
   \begin{pmatrix}
    G_{i_{n+1}, j_1}^{(0)} & G_{i_{n+1},j_2}^{(0)} & \cdots & G_{i_{n+1},j_n}^{(0)} \\
    G_{i_2, j_1}^{(0)}     & G_{i_2,j_2}^{(0)}     & \cdots & G_{i_2,j_n}^{(0)}     \\
    \vdots                 & \vdots 		   & \ddots & \vdots 		    \\
    G_{i_n, j_1}^{(0)}     & G_{i_n,j_2}^{(0)}     & \cdots & G_{i_n,j_n}^{(0)}
   \end{pmatrix}  \nonumber \\
 &- G_{i_2, j_{n+1}}^{(0)} \cdot\det
   \begin{pmatrix}
    G_{i_1, j_1}^{(0)} & G_{i_1,j_2}^{(0)} & \cdots & G_{i_1,j_n}^{(0)} \\
    G_{i_{n+1}, j_1}^{(0)}     & G_{i_{n+1},j_2}^{(0)}     & \cdots & G_{i_{n+1},j_n}^{(0)}     \\
    \vdots                 & \vdots 		   & \ddots & \vdots 		    \\
    G_{i_n, j_1}^{(0)}     & G_{i_n,j_2}^{(0)}     & \cdots & G_{i_n,j_n}^{(0)}
   \end{pmatrix}
 - \cdots \nonumber \\
 &- G_{i_n, j_{n+1}}^{(0)} \cdot\det
   \begin{pmatrix}
    G_{i_1, j_1}^{(0)} & G_{i_1,j_2}^{(0)} & \cdots & G_{i_1,j_n}^{(0)} \\
        \vdots                 & \vdots 		   & \ddots & \vdots 		    \\
    G_{i_{n-1}, j_1}^{(0)}     & G_{i_{n-1},j_2}^{(0)}     & \cdots & G_{i_{n-1},j_n}^{(0)}     \\
    G_{i_{n+1}, j_1}^{(0)}     & G_{i_{n+1},j_2}^{(0)}     & \cdots & G_{i_{n+1},j_n}^{(0)}
   \end{pmatrix} \nonumber \\
 &+ G_{i_{n+1}, j_{n+1}}^{(0)} \cdot\det
   \begin{pmatrix}
    G_{i_1, j_1}^{(0)} & G_{i_1,j_2}^{(0)} & \cdots & G_{i_1,j_n}^{(0)} \\
    G_{i_2, j_1}^{(0)}     & G_{i_2,j_2}^{(0)}     & \cdots & G_{i_2,j_n}^{(0)} \\
	\vdots                 & \vdots 		   & \ddots & \vdots 		    \\
    G_{i_n, j_1}^{(0)}     & G_{i_n,j_2}^{(0)}     & \cdots & G_{i_n,j_n}^{(0)}
   \end{pmatrix}. \label{eq:induction_step} 
\end{align} 
By rearranging rows in \eqref{eq:induction_step}, it can be concluded that \eqref{eq:induction_step}
is the expansion of the $(n+1) \times (n+1)$ determinant 
\begin{equation}
 \det 
 \begin{pmatrix}
  G_{i_1,j_1}^{(0)} & G_{i_1,j_2}^{(0)} & \cdots & G_{i_1,j_{n+1}}^{(0)} \\
  G_{i_2,j_1}^{(0)} & G_{i_2,j_2}^{(0)} & \cdots & G_{i_2,j_{n+1}}^{(0)} \\
  \vdots            &   \vdots          & \ddots & \vdots          \\
  G_{i_{n+1},j_1}^{(0)}   & G_{i_{n+1},j_2}^{(0)}   & \cdots & G_{i_{n+1}, j_{n+1}}^{(0)}
 \end{pmatrix}
\end{equation}
along the last column according to Laplace's formula. This completes the inductive proof.
Except for the last determinant in Eq.~\eqref{eq:induction_step}, row exchanges are necessary to obtain the 
correct submatrix structure. 
In the determinant accompanying the single-particle Green's 
function $G_{i_k, j_{n+1}}^{(0)}$ in Eq.~\eqref{eq:induction_step}
we need to perform $(n-k)$ row exchanges which results in 
a factor $(-1)^{n-k}$ such that the total sign is $(-1)^{n-k+1}$,
which is identical to the alternating factor $(-1)^{n+1+k}$ coming from Laplace's formula.

Eq.~(5) differs from Eq.~\eqref{eq:Wick_determinant} in that instead of pairings $(\hat{c}_{i_k} \hat{c}_{j_k}^{\dagger})$  
there are projectors of the form
\begin{equation}
 \Pi_k = n_k \hat{c}_{j_k}^{\dagger} \hat{c}_{j_k} + (1 - n_k) \hat{c}_{j_k} \hat{c}_{j_k}^{\dagger}.
 \label{eq:projector_k}
\end{equation}
Depending on whether $n_k=0$ or $n_k=1$, only one of either terms in \eqref{eq:projector_k} applies. 
Let us replace for the moment only the $k$-th pairing $(\hat{c}_{i_k} \hat{c}_{j_k}^{\dagger})$ in Eq.~\eqref{eq:Wick_determinant} by $\Pi_k$. 
If $n_k = 0$, the resulting expression is covered by Eq.~\eqref{eq:Wick_determinant}. The case $n_k = 1$ is different since 
$\hat{c}_{j_k}^{\dagger}$ is to the right of $\hat{c}_{j_k}$. Using $\hat{c}_{j_k}^{\dagger} \hat{c}_{j_k} = 1 - \hat{c}_{j_k} \hat{c}_{j_k}^{\dagger}$, one obtains
\begin{equation}
\langle (\hat{c}_{i_1} \hat{c}_{j_1}^{\dagger} \cdots \Pi_k \cdots (\hat{c}_{i_n} \hat{c}_{j_n}^{\dagger}) \rangle_{0} 
= (-1)^{n_k}  n_k  \left[\det\left( G^{(0)}_{I \backslash j_k, J \backslash j_k} \right) - \det(G^{(0)}_{I,J}) \right]
\label{eq:commute_annihilator_creator}
\end{equation}
To make the connection with Eq.~(5) we consider the matrix 
\begin{equation}
 \left( \tilde{G}^{(0)} \right)_{ij} = \left( G^{(0)} \right)_{ij} - n_k \delta_{i,j_k} \delta_{j, j_k}
 \label{eq:Gtilde}
\end{equation}
and develop its determinant with respect to the $k$-th column:
\begin{align}
 \det(\tilde{G}) &= \sum_{l=1}^{n} (-1)^{l + k} \left( \tilde{G^{(0)}} \right)_{i_l, j_k}  \left[ \tilde{G^{(0)}} \right]_{I \backslash i_l, J\backslash j_k} \\
 &= \sum_{l=1}^{n} (-1)^{l+k} \left( G^{(0)} \right)_{i_l, j_k} \left[ G^{(0)} \right]_{I \backslash i_l, J \backslash j_k} - n_k (-1)^{2k} \left[ G^{(0)} \right]_{I \backslash j_k, J \backslash j_k} \label{eq:undo_Laplace}
\end{align}
Here, angular braces $( \cdot )$ denote matrix elements and square brackets $[ \cdot ]$ denote the inclusive definition of the minor. 
Note that the minors $\left[ \tilde{G^{(0)}} \right]_{I \backslash i_l, J \backslash j_k} = \left[ G^{(0)} \right]_{I\backslash i_l, J \backslash j_k}$ 
since $\tilde{G^{(0)}}$ and $G^{(0)}$ only differ in the element $(j_k, j_k)$ which is excluded from the minors.
Now, one can recognize in the first sum Eq.~\eqref{eq:undo_Laplace} the Laplace expansion of $G^{(0)}$ with respect to the $k$-th column and 
undo it again to recover $\det(G^{(0)})$:
\begin{equation}
 \det\left(\tilde{G^{(0)}}\right) = \det\left( G^{(0)} \right) - n_k \left[ G^{(0)} \right]_{I \backslash j_k, J \backslash j_k}.
 \label{eq:undo_Laplace_expansion}
\end{equation}
Eq.~\eqref{eq:undo_Laplace_expansion} is identical to Eq.~\eqref{eq:commute_annihilator_creator}, which proves that 
\begin{align}
\langle (\hat{c}_{i_1} \hat{c}_{j_1}^{\dagger}) \cdots \Pi_k \cdots (\hat{c}_{i_n} \hat{c}_{j_n}^{\dagger}) \rangle_{0} = \det 
\begin{pmatrix}
G^{(0)}_{i_1, j_1} & G^{(0)}_{i_1, j_2} & \cdots & \cdots & G^{(0)}_{i1, j_n} \\
G^{(0)}_{i_2, j_1} & G^{(0)}_{i_2, j_2} & \cdots & \cdots & G^{(0)}_{i2, j_n} \\
\vdots       & \vdots       & \ddots & \cdots &  \vdots     \\
\vdots       & \vdots       & G^{(0)}_{j_k, j_k} - n_k & \cdots &  \vdots       \\
\vdots       & \vdots       & \vdots 	         & \ddots &  \vdots       \\
G^{(0)}_{i_n, j_1} & G^{(0)}_{i_n, j_2} & \cdots & \cdots & G^{(0)}_{i_n, j_n} \\
\end{pmatrix}.
\label{eq:proof_one_projector}
\end{align}
Repeatedly replacing each pairing $(\hat{c}_{i_l} \hat{c}_{j_l}^{\dagger})$ in Eq.~\eqref{eq:Wick_determinant} by a projector 
$\Pi_l$ of the form \eqref{eq:projector_k} and repeating the derivation from \eqref{eq:Gtilde} to \eqref{eq:proof_one_projector}, 
with a Laplace expansion carried out with respect to the $l$-th column, completes the proof of Eq.~(5).

\section{Exploitation of block matrix structure}
\label{app:block_determinant}

Eq.~(5) implies that the expressions for the joint quasiprobability 
distributions of successive numbers of components 
are related by a block matrix structure. 
Using the formula for the determinant of a block matrix and noticing that $G(y,y)$ is just a number:
\begin{align}
 p(x_1, x_2, \ldots, x_{k-1}; y) &= \frac{1}{Z} \det 
 \left(
 \begin{array}{c|c}
 \raisebox{-35pt}{{\huge\mbox{{$X_{k-1}$}}}} & G(x_1, y) \\
                                       & G(x_2, y) \\
				       & \vdots    \\ 
				       & G(x_{k-1}, y) \\ \hline
 G(y,x_1) \, G(y,x_2) \, \cdots \, G(y,x_{k-1}) & G(y,y)  
 \end{array}
 \right)\\
 &= \frac{1}{Z} \det(X_{k-1}) \left[G(y,y) - \sum_{i,j=1}^{k-1} G(y,x_i) \left[ X_{k-1}^{-1}\right]_{i,j} G(x_j,y) \right].
\end{align}

Given that $X_{k-1}$ is itself a block matrix 
\begin{align}
 X_{k-1} &=
 \left(
 \begin{array}{c|c}
 \raisebox{-35pt}{{\huge\mbox{{$X_{k-2}$}}}} 			  & G(x_1, x_{k-1}) \\
								  & G(x_2, x_{k-1}) \\
								  & \vdots    \\ 
								  & G(x_{k-2}, x_{k-1}) \\ \hline
 G(x_{k-1},x_1) \, G(x_{k-1},x_2) \, \cdots \, G(x_{k-1},x_{k-2}) & G(x_{k-1},x_{k-1})  
 \end{array}
 \right)
\end{align} 
with $G(x_{k-1},x_{k-1})$ just a number and assuming that the inverse $X_{k-2}^{-1}$ is already known, one 
can make use of the formula for the inversion of a block matrix to compute the inverse of $X_{k-1}$ in 
an economical way.
We define
\begin{equation}
 g \equiv G(x_{k-1}, x_{k-1}) - \sum_{i,j=1}^{k-2} G(x_{k-1}, x_i) \left[ X_{k-2}^{-1} \right]_{i,j} G(x_j, x_{k-1}) 
\end{equation}
and recognize that
\begin{equation}
 g = \frac{p(x_1, x_2, \ldots, x_{k-1})}{p(x_1, x_2, \ldots, x_{k-2})} = p(x_{k-1}| x_{k-2}, \ldots, x_2, x_1),
\end{equation}
which means that we have computed $g$ already previously when sampling the $(k-1)$-th component.

Using the formula for the inverse of a block matrix
\begin{equation}
 X_{k-1}^{-1} = \left(
 \begin{array}{c|c}
   X_{k-2}^{-1} + g^{-1} \vec{u} \otimes \vec{v}^{T}    &    - g^{-1} \vec{u} \\  \hline
                - g^{-1} \vec{v}^{T}                    &      g^{-1}
 \end{array}
 \right),
\end{equation}
where 
\begin{align}
 [\vec{u}]_i &= \sum_{j=1}^{k-2} \left[ X_{k-2}^{-1} \right]_{ij} G(x_j, x_{k-1}), \\
 [\vec{v}^{T}]_j &= \sum_{i=1}^{k-2} \left[ X_{k-2}^{-1}\right]_{ij} G(x_{k-1}, x_i),
\end{align}
and 
\begin{equation}
 \left[\vec{u} \otimes \vec{v}^{T} \right]_{ij} = [\vec{u}]_{i} [\vec{v}^{T}]_{j}.
\end{equation}

Thus, the update $X_{k-2}^{-1} \rightarrow X_{k-1}^{-1}$ requires the computation of $\vec{u}$, $\vec{v}^{T}$,
and the exterior product $\vec{u} \otimes \vec{v}^{T}$, which is of order $\mathcal{O}((k-2)^2)$.
It is easy to see that the sampling of $N_p$ particle positions requires $\mathcal{O}(\sum_{i=1}^{N_p} i^2) = \mathcal{O}(N_p^3)$
floating point operations. It is not necessary to compute any inverse or determinant from scratch.

\section{Sign problem for nested componentwise direct sampling at and away from half filling}
\label{app:sign_problem}

The severity of the conventional sign problem in the determinantal QMC algorithm is measured by the average sign
of the Monte Carlo weight 
\begin{equation}
\langle s \rangle = \frac{1}{N_{\text{HS samples}}} \sum_{\{ \vecbf{s}\}} \text{sign}(w_{\{ \vecbf{s} \}}^{\uparrow}  w_{\{ \vecbf{s} \}}^{\downarrow}),
\end{equation}
where the sum is over auxiliary field configurations of the Hubbard-Stratonovich samples.
\begin{figure}
\includegraphics[width=0.75\linewidth]{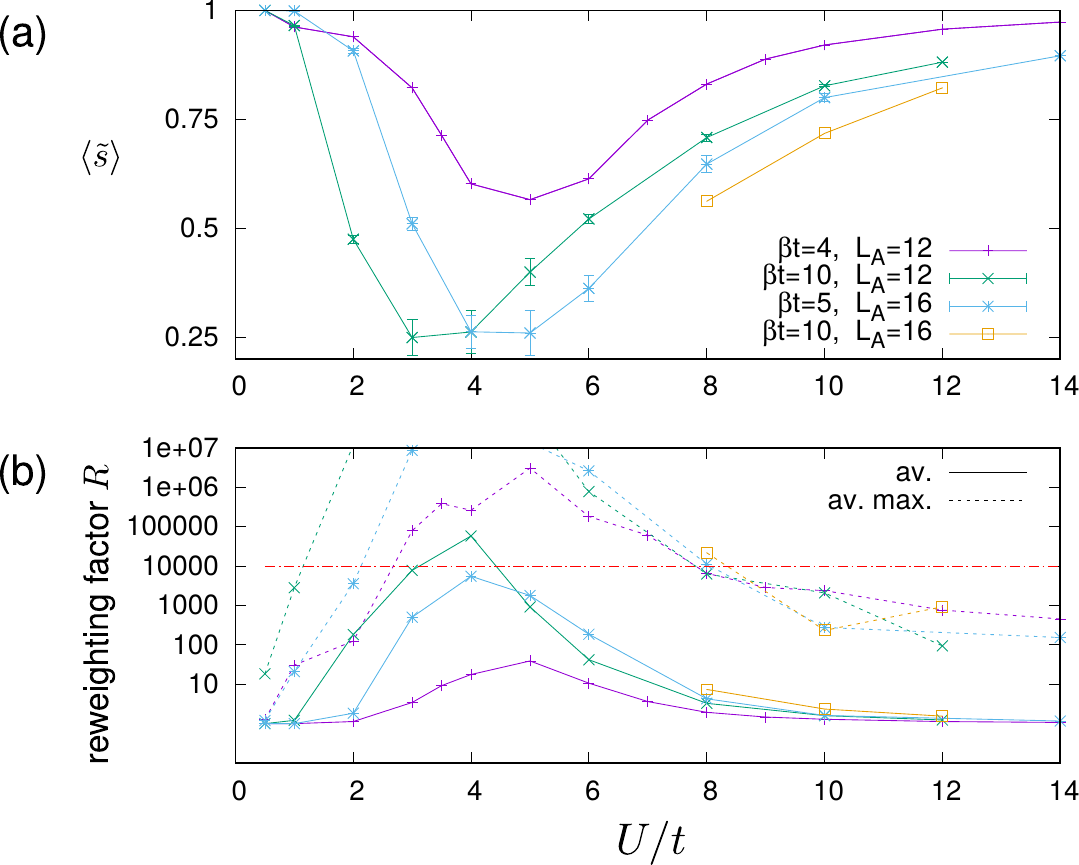}
\caption{Sampling sign problem at half filling. The probe area is the total $L \times L$ square system ($L=L_A$).
(a) Average sign of a pseudo-snapshot. (b) Average (unsigned) reweighting factor (full lines) and average maximum reweighting factor (dashed lines).
The red dashed-dotted line indicates a rough estimate of the threshold of maximum reweighting factors below which numerically exact results 
can be obtained with modest computational effort.}
\label{fig:av_sign_halffilling}
\end{figure}
\begin{figure}
\includegraphics[width=0.75\linewidth]{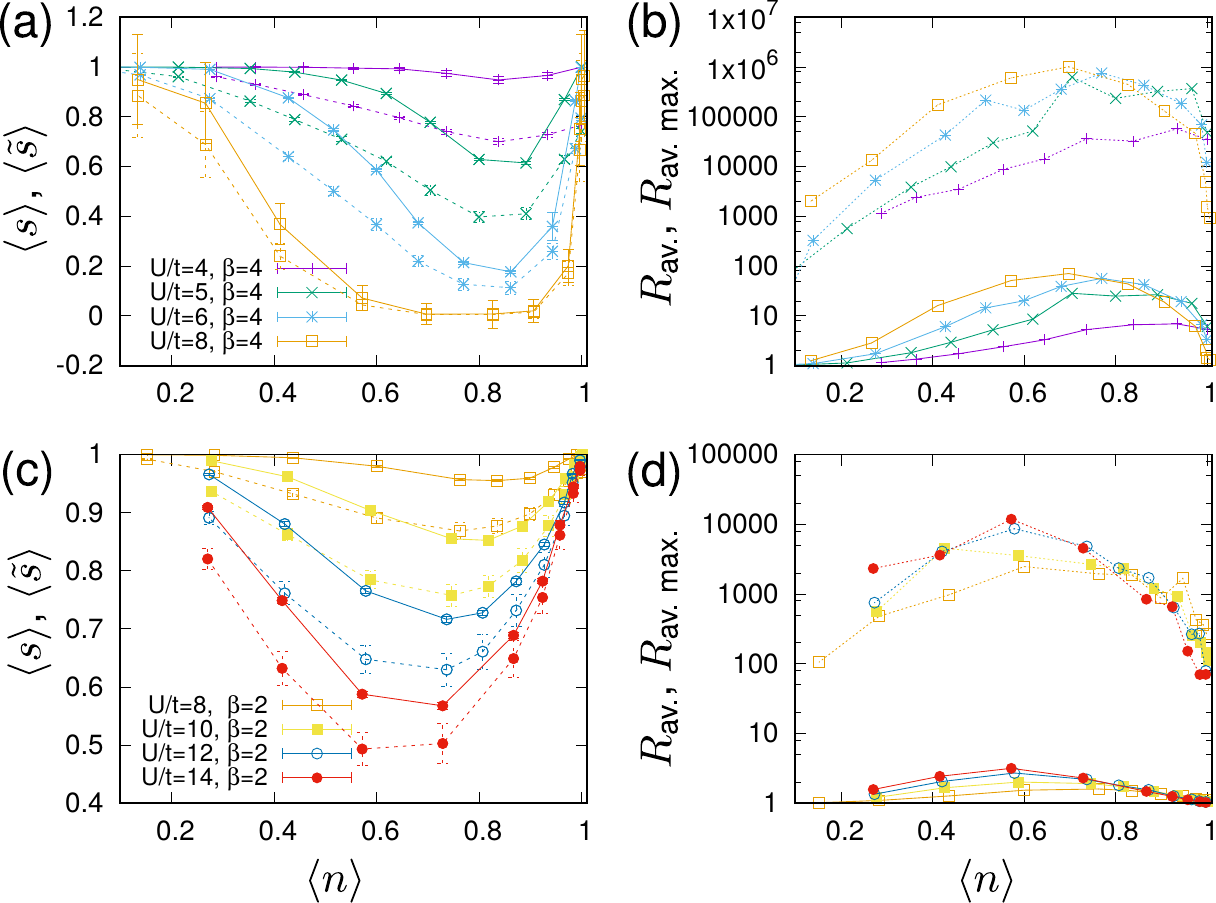}
\caption{Conventional sign problem and sampling sign problem 
away from half filling: dependence on $U/t$. (a,c) Average sign ($\langle s \rangle$, full lines) and average sampling sign ($\langle \tilde{s} \rangle$, dashed lines);
(b,d) average reweighting factor (full lines) and average maximum reweighting factor (dotted lines).
The inverse temperature $\beta$ is in units of $1/t$.
The probe area is equal to the system size with $L_A=L=8$.}
\label{fig:av_sign_doped_U}
\end{figure} 
\begin{figure}
 \includegraphics[width=0.75\linewidth]{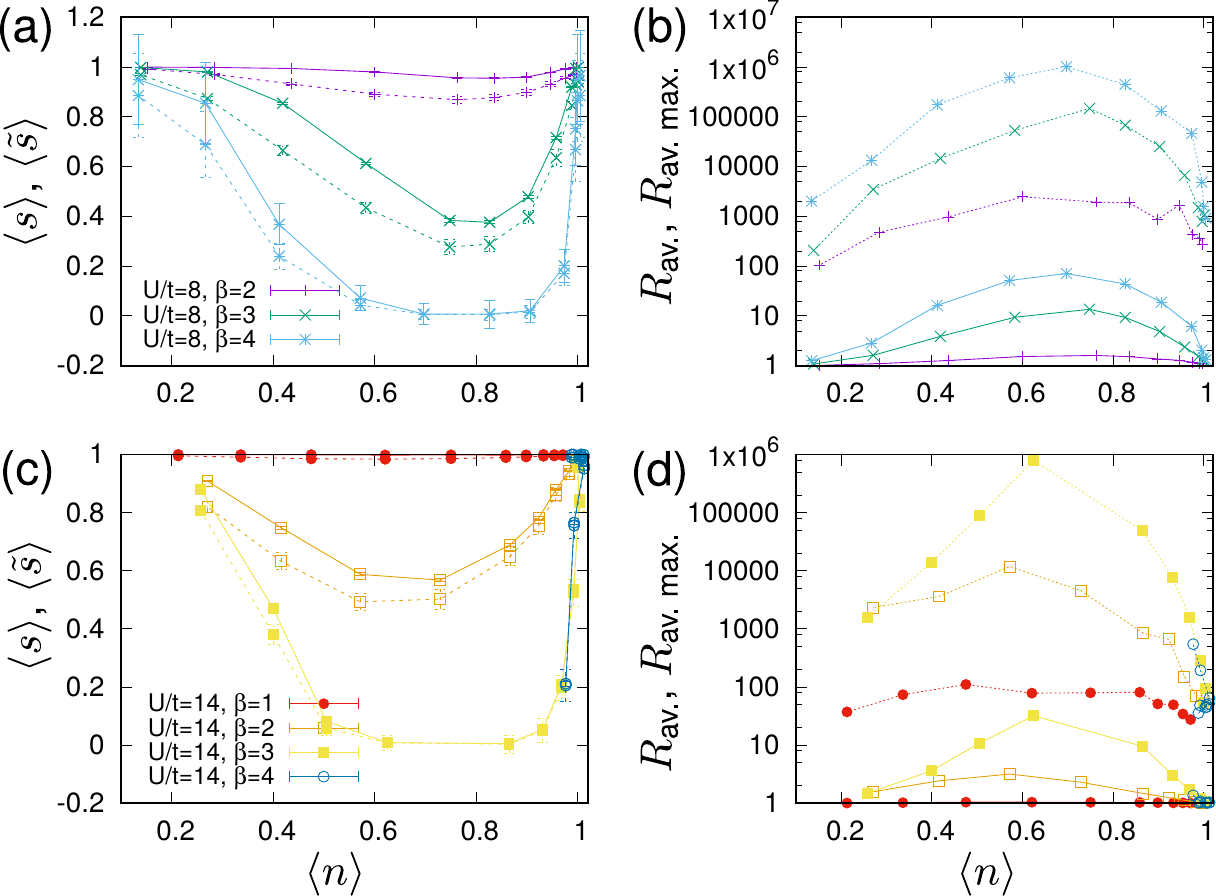}
 \caption{Conventional sign problem  and sampling sign problem
 away from half filling: dependence on inverse temperature $\beta t$. System size $L_A=L=8$.}
 \label{fig:av_sign_doped_temp}
\end{figure}
To quantify the ``sampling sign problem'' in the nested componentwise sampling algorithm we introduce 
the average sign of a snapshot, $\langle \tilde{s} \rangle$, which is given by
\begin{equation}
\langle \tilde{s} \rangle = \frac{1}{M}\sum_{\{ \vecbf{s}\}} \text{sign}(w_{\{ \vecbf{s} \}}^{\uparrow}  w_{\{ \vecbf{s} \}}^{\downarrow}) 
\sum_{i=1}^{N_{\text{snapshots per HS}}} \prod_{\sigma = \uparrow, \downarrow} \prod_{k=1}^{D} \text{sign}(p^{(i)}_{\{ \vecbf{s} \}}(n^{\sigma}_k)),
\label{eq:average_sampling_sign}
\end{equation}
where $M = N_{\text{HS samples}} \times N_{\text{snapshots per HS}}$ is the total number of generated snapshots.
Here, it is understood that snapshots drawn from the same HS sample, the number of which is $N_{\text{snapshots per HS}}$,
are multiplied by the sign of the 
corresponding Monte Carlo weight $w_{\{ \vecbf{s} \}}^{\uparrow}  w_{\{ \vecbf{s} \}}^{\downarrow}$,
in case that there is already a conventional sign problem at the level of the determinantal QMC algorithm. 
Note that within one HS sample, $\{ \vecbf{s} \}$, snapshots for spin-$\uparrow$ and spin-$\downarrow$
can be paired up arbitrarily into a full snapshot due to the statistical independence of the two spin species.

Furthermore, one can define the average (unsigned) reweighting factor 
\begin{equation}
R_{\text{av.}} = \frac{1}{M} \sum_{\{ \vecbf{s} \}} \sum_{i=1}^{N_{\text{snapshots per HS}}} 
\prod_{\sigma = \uparrow, \downarrow} \prod_{k=1}^{D} \mathcal{N}^{{\sigma}, (i)}_{k, \{ \vecbf{s}\}}
\end{equation}
and the average maximum reweighting factor, averaged over independent Markov chains, 
\begin{equation}
  R_{\text{av. max.}} = \frac{1}{N_{\text{Markov chains}}} \sum_{m=1}^{N_{\text{Markov chains}}} \max_{\substack{\text{snapshots } i \\ \text{from m-th Markov chain}}}\left( |R_i| \right).
\end{equation}
The average maximum reweighting $R_{\text{av. max.}}$ is the most relevant indicator of the severity of the sign problem
as it quantifies the ''inflation`` of value of an individual snapshot. Because the reweighting factor can fluctuate 
over several orders of magnitude, the average sign alone is not sufficient for a characterization of the ''sampling sign problem``. 
How the sampling sign and the reweighting factor depend on doping, inverse temperature $\beta t$
and interaction strength $U/t$ is shown in
Figs. \ref{fig:av_sign_halffilling}, \ref{fig:av_sign_doped_U}, and \ref{fig:av_sign_doped_temp} .

The snapshots generated in quantum gas microscope experiments originate from 
independent experimental runs whereas the pseudo-snapshots of the nested componentwise sampling 
are affected by the autocorrelation time inherent in the sampling of Hubbard-Stratonovich field 
configurations via the standard determinantal QMC algorithm. 
For large $U/t$, pseudo-snapshots taken from the same HS sample (and at the same imaginary time slice), differ mostly in the positions 
of holes and doubly occupied sites, while the spin background stays largely fixed. 

Additionally, pseudo-snapshots come with a sign and reweighting factor 
due to the non-Hermiticity of the pseudo-density operator within each HS sample. 
Therefore, experimental snapshots and pseudo-snapshots are not directly comparable. 
An important question is how many pseudo-snapshots $M$ are required to obtain a comparable precision 
of measurement quantities as from $M^{(\text{exp.})}$ experimental snapshots. 
Taking into account the average maximum reweighting factor, the effective number of pseudo-snapshots $M^{\prime}$
that is equivalent to $M^{(exp)}$ can be roughly estimated as 
\begin{equation}
 M^{\prime} = M / (\tau_{\text{AC}}^{\text{DQMC}} \times R_{\text{av.max.}}) \leftrightarrow  M^{(\text{exp.})}.
\end{equation}
Here, $\tau_{\text{AC}}^{\text{DQMC}} $ is some measure of the autocorrelation of the Markov chain generated 
by the standard determinantal QMC algorithm. 

The ''sampling sign`` deteriorates exponentially with the probe area $N_{A} = L_A^2$. Yet, an extent of the probe area $L_A \sim \xi(T/t, U/t, L)$,
where $\xi$ is the correlation length, is often sufficient for meaningful simulations. 

\section{Code verification}
\label{app:benchmark_pentagon}
\begin{figure}[h!]
\includegraphics[width=0.65\linewidth]{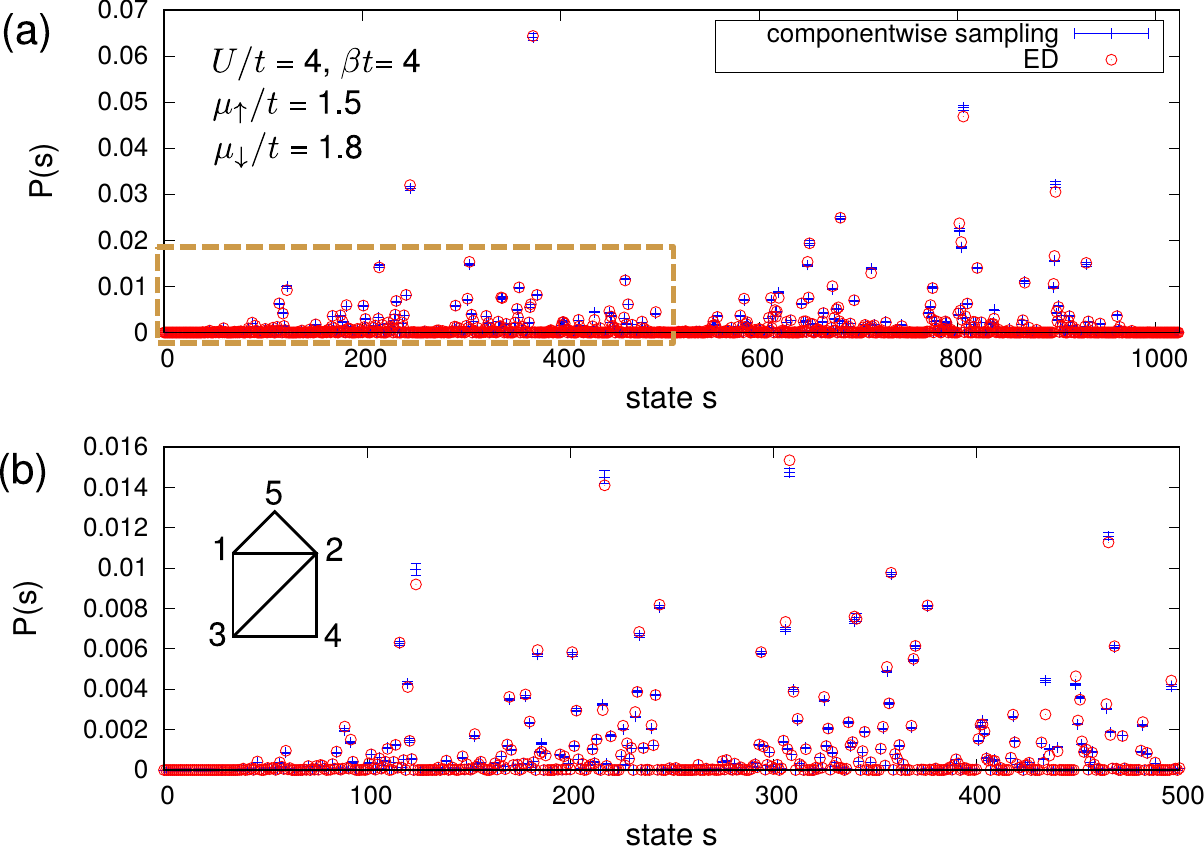}
\caption{Probabilities $P(s)$ of all microstates $s$ in the occupation number basis.
Comparison with exact diagonalization (ED) for the parameters shown in the upper panel and for the hopping matrix of Eq.~\eqref{eq:benchmark_hopping}. 
The total number of Fock samples is $10^7$, generated 
by 56 independent Markov chains.}
\label{fig:benchmark_5sites}
\end{figure}

\begin{figure}[h!]
\includegraphics[width=1.0\linewidth]{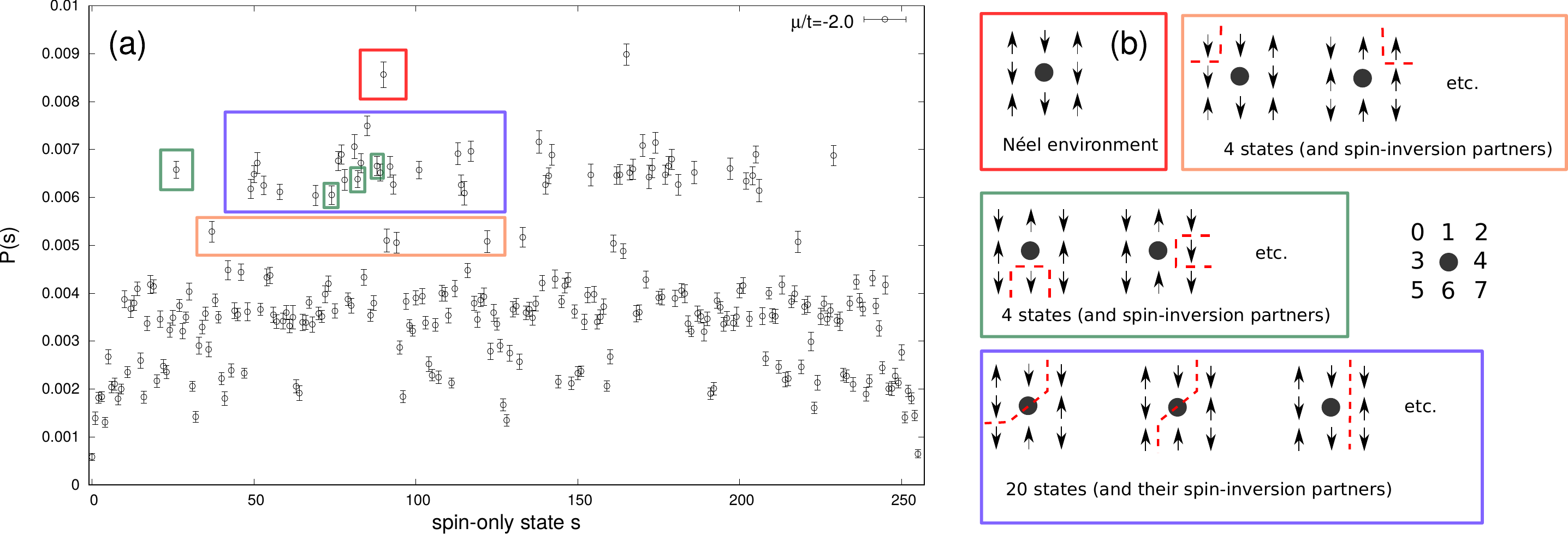}
\caption{(a) Probabilities of spin-only states around an empty site as obtained from reweighted pseudo-snapshots
in the reference frame of the hole. Parameters: $U/t=12, \mu/t = -2, \beta t = 2.5$, $L \times L$
square lattice with $L=8$. 
For this system size, the average number of holes and doubles $\langle N_h \rangle \approx 1.50$ and $\langle N_d \rangle \approx 1.31$
so that the average number of excess holes is $\langle N_h \rangle - \langle N_d \rangle \approx 0.19$. 
With 56 independent Markov chains a total number of $70 \times 10^6$ occupation number samples of 
the entire system was obtained. 
In the reference frame of the hole, those states around the hole position are filtered out
which do not contain charge fluctuations (``spin-only states''). 
Per Hubbard-Stratonovich configuration, 200 direct sampling steps were performed during the Monte Carlo sweeps 
at different imaginary time slices. Trotter discretization $\Delta \tau t= 0.02$.
(b) Illustration of the most important spin-only states around an isolated hole which are marked in panel (a).
Dashed lines delineate different antiferromagnetic domains. }
\label{fig:hole_spingb_environment}
\end{figure}

For benchmarking purposes an irregular model instance of the Hubbard model on five sites (see inset in Fig.~\ref{fig:benchmark_5sites}(b)) is chosen
\begin{equation}
 H = -\sum_{\langle i,j, \rangle, \sigma=\uparrow, \downarrow} t^{\sigma}_{ij}\left( c_{i,\sigma}^{\dagger} c_{j,\sigma} + h.c. \right) 
 + U \sum_{i=1}^{5} n_{i,\uparrow} n_{i,\downarrow}  - \sum_{\sigma = \uparrow, \downarrow}\mu_{\sigma} \sum_{i=1}^{5} n_{i,\sigma},
\end{equation}
which breaks translational, point group and spin rotational 
symmetry. The hopping matrix is identical for both spin species and reads
\begin{equation}
 [t^{\sigma}]_{i,j} = t
\begin{pmatrix}
0 & 0.7 & 1.1 & 0 & 0.8 \\
0.7 & 0 & 1.05 & 0.9 & 1.2 \\
1.1 & 1.05 & 0 & 1.0 & 0 \\
0 & 0.9 & 1.0 & 0 & 0 \\
0.8 & 1.2 & 0 & 0 & 0 \\
\end{pmatrix}
\label{eq:benchmark_hopping}
\end{equation}.
The onsite repulsion is $U/t=4$ and the chemical potential for spin up and down is 
$\mu_{\uparrow}/t=1.5$ and $\mu_{\downarrow}/t=1.8$, respectively; the inverse temperature is 
$\beta t = 4$ discretized into $N_{\tau} = 256$ Trotter time slices with $\Delta \tau t = \beta t / N_{\tau} = 1/64$.
Fig.~\ref{fig:benchmark_5sites} compares the probabilities $P(s)$ of all microstates $s$ in the occupation number basis 
with results from exact diagonalization. An enlarged view of Fig.~\ref{fig:benchmark_5sites}(a) is shown in Fig.~\ref{fig:benchmark_5sites}(b).
The occupation number state $\vec{n} = [n_{1, \uparrow} \ldots n_{5, \uparrow}; n_{1, \downarrow} \ldots n_{5, \downarrow}]$ with $n_{i,\sigma} \equiv n_{\alpha} \in \{0, 1\}$
is interpreted as a bitstring and is represented by the integer $s = \sum_{\alpha = 0}^{2 N_{\text{sites}}-1} n_{\alpha} 2^{\alpha}$.
With the knowledge of the eigenenergies $E_i$ and eigenvectors $|\phi_i \rangle$ of the Hamiltonian $H$ the probability of 
an occupation number state $s$ is
\begin{equation}
 P(s) = \sum_{i=1}^{\text{dim}(H)} \frac{e^{-\beta E_i}}{Z} |\langle s | \varphi_i \rangle |^2
\end{equation}
with $Z = \sum_{i=1}^{\text{dim}(H)} e^{-\beta E_i}$.
In order to demonstrate that the nested componentwise direct sampling method yields consistent microscopic states also for much larger system sizes, 
Fig.~\ref{fig:hole_spingb_environment} shows the probabilities of spin-only states on the eight sites around an isolated, mobile hole
for a system of $10 \times 10$ lattice sites at small doping. Only isolated holes, i.e. those surrounded exclusively by singly occupied sites, are considered 
(see Eq.~(13) of the main text).
With the labelling of sites around a hole as shown in the right panel of Fig.~\ref{fig:hole_spingb_environment}, the spin environment is characterized 
by the state $\vec{\sigma} = [\sigma_{0} \sigma_{1} \ldots \sigma_{7}]$ where $\sigma_{p}=1$ 
if there is a spin-$\uparrow$ at position $p$ around the hole and $\sigma_{p}=0$ if it is a spin-$\downarrow$. 
With this convention, the state index in Fig.~\ref{fig:hole_spingb_environment} is given as $s = \sum_{p = 0}^{7} \sigma_{p} 2^{p}$. 
Fig.~\ref{fig:hole_spingb_environment} shows that there is a hierarchy of groups of states (right panel of Fig.~\ref{fig:hole_spingb_environment}).
Furthermore, states related by symmetry, which are grouped together in coloured boxes in Fig.~\ref{fig:hole_spingb_environment},
appear with approximately the same probability.
This is not a built-in feature of the nested componentwise sampling algorithm and thus 
provides strong evidence that all relevant occupation number states are sampled with their correct probabilities. 

\section{FCS of number of doublons and holes}
\label{app:FCS_doublons_holes}

\begin{figure}
\includegraphics[width=0.5\linewidth]{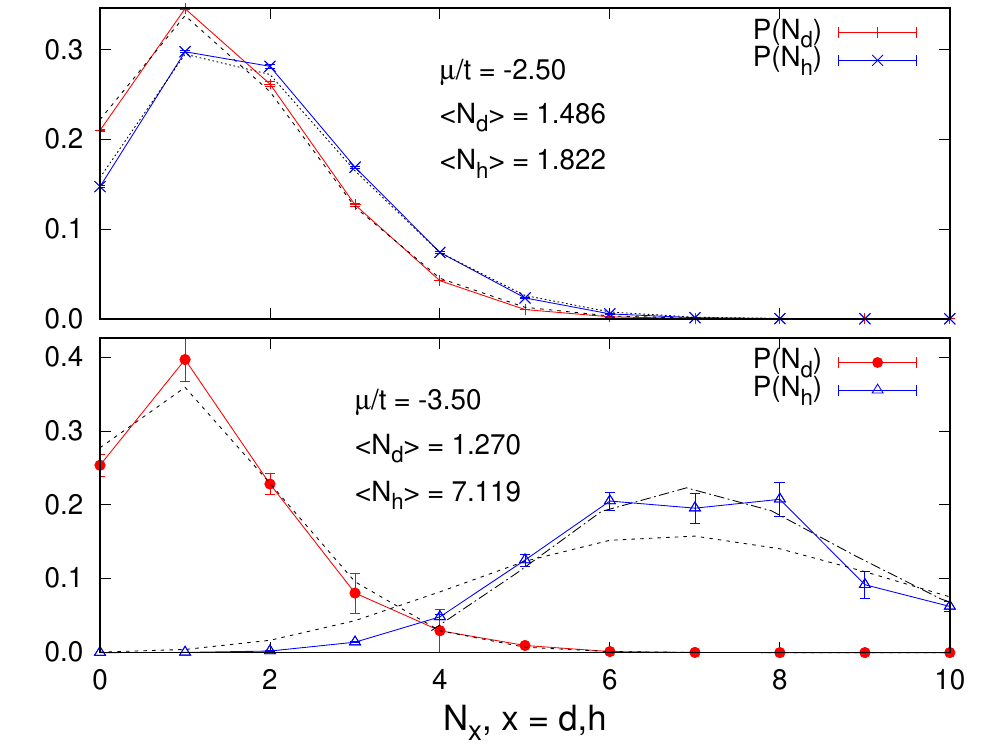}
\caption{FCS of the total number of doubly occupied sites $\langle N_d \rangle$ and holes $\langle N_h \rangle$, obtained from reweighted pseudo-snapshots
for different levels of hole doping (top and bottom panel). Dashed and dotted lines are binomial distributions; the dashed-dotted line 
is a ``shifted'' binomial distribution.
Parameters: $U/t=12$, $\beta t = 4, L=8$.} 
\label{fig:FCS_doublons_holes}
\end{figure}

Fig.~\ref{fig:FCS_doublons_holes} displays the doping dependence of the FCS of the number of doubly occupied sites 
$N_{d} = \sum_{i=1}^{N_{A}} n_{i,\uparrow} n_{i,\downarrow}$  
and holes $N_h = \sum_{i=1}^{N_{A}} \left(1 - n_{i,\uparrow} \right) \left(1 - n_{i,\downarrow} \right)$
for $U/t=12, \beta t = 4$ on an $L \times L$ system with $L=L_A=8$. 
For small doping values, the FCS are well described by 
binomial distributions (dashed and dotted lines) which are generated by populating 
lattice sites independently with doublons (holes) with probability $p_d = \langle n_{\uparrow} n_{\downarrow} \rangle$ 
($p_h = \langle (1 - n_{\uparrow}) (1 - n_{\downarrow} \rangle) \rangle$).
according to the distribution 
$\mathcal{B}_{n,p}(x) = \begin{pmatrix} n \\ x\end{pmatrix} p^{x} (1-p)^{n-x}$ with $n=N_{A}$ and \mbox{$p \in \{p_d, p_h \}$}. 
In the strongly doped case (lower panel), the distribution of the number of holes can be modelled accurately 
by a ``shifted'' binomial distribution (dashed-dotted line), which 
is obtained by populating the lattice with holes arising from doublon-hole fluctuations according to a binomial distribution 
with parameter $p=p_d$ (since the number of doublon-hole pairs is approximately equal to the number of doublons) 
which is then shifted so that the distribution is centered about its mean.
Of course, for a large probe area and away from a critical point, the distribution of holes (doublons) should 
approach a Gaussian distribution which could be characterized 
by computing the mean and variance directly.

\section{Additional data for the polaron problem}
\label{app:additional_data}

\begin{figure}[b!]
\includegraphics[width=0.95\textwidth]{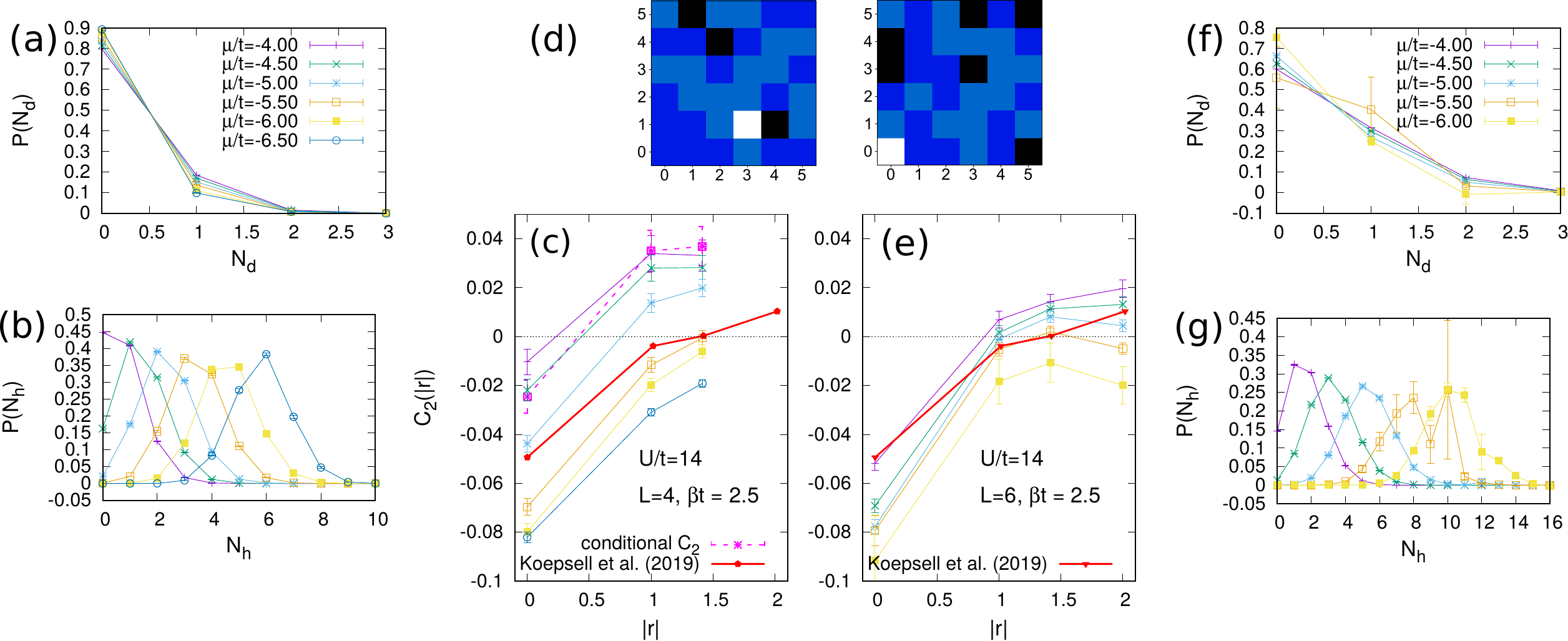}
\caption{The three-point spin-charge correlation function $C_{|\vecbf{d}|=2}(|\vecbf{r}|)$ of Eq.~\eqref{eq:cond_spin_charge_no_projector}
computed on a $4 \times 4$ system (a-c) as well as a $6 \times 6$ 
system (d-g) with periodic boundary conditions and with parameters $U/t=14, \beta t = 2.5, \mu=-4.0, -4.5, \ldots, -6.5$.
The dashed magenta line in (c) [denoted ``conditional $C_2$''] shows the correlation function  $C_{|\vecbf{d}|=2}(|\vecbf{r}|)$ calculated for $\mu/t=-4.0$ according to 
Eq.~\eqref{eq:third_order_corr} in the main text, in which only those snapshots are counted where the hole is surrounded on eight sites 
by spin-only states (i.e. it is not adjacent to a doublon or a hole and can be regarded as an isolated dopant). 
The experimental data of Ref.~\cite{Koepsell2019}
are also shown for comparison. There, the system size is approximately $4 \times 6$ to $6 \times 6$ lattice sites (slightly inhomogeneous with open boundary conditions), interactions $U/t \approx 14$
and stated temperature $T = 1.4 J$, which corresponds to $T/t \approx 0.4$ (i.e. $\beta t = 2.5$) for $J= 4 t^2 / U$. 
In the quantum gas microscope experiment, there are on average $1.95(1)$ dopants present 
in each experimental realization (doublon-doped instead of hole-doped) \cite{Koepsell2019}. 
For our setting on a $4 \times 4$ system at $\mu/t=-4.0$, on the other hand, the average number of holes (hole-doped) 
is $\langle N_h \rangle = 0.72$, which is small enough for measuring the spin-environment of an \emph{isolated} dopant.
(d) Randomly chosen pseudo-snapshots for $\mu/t=-4$ (left) and $\mu/t=-5$ (right).}
\label{fig:no_polaron}
\end{figure}
\begin{figure}
\includegraphics[width=0.5\textwidth]{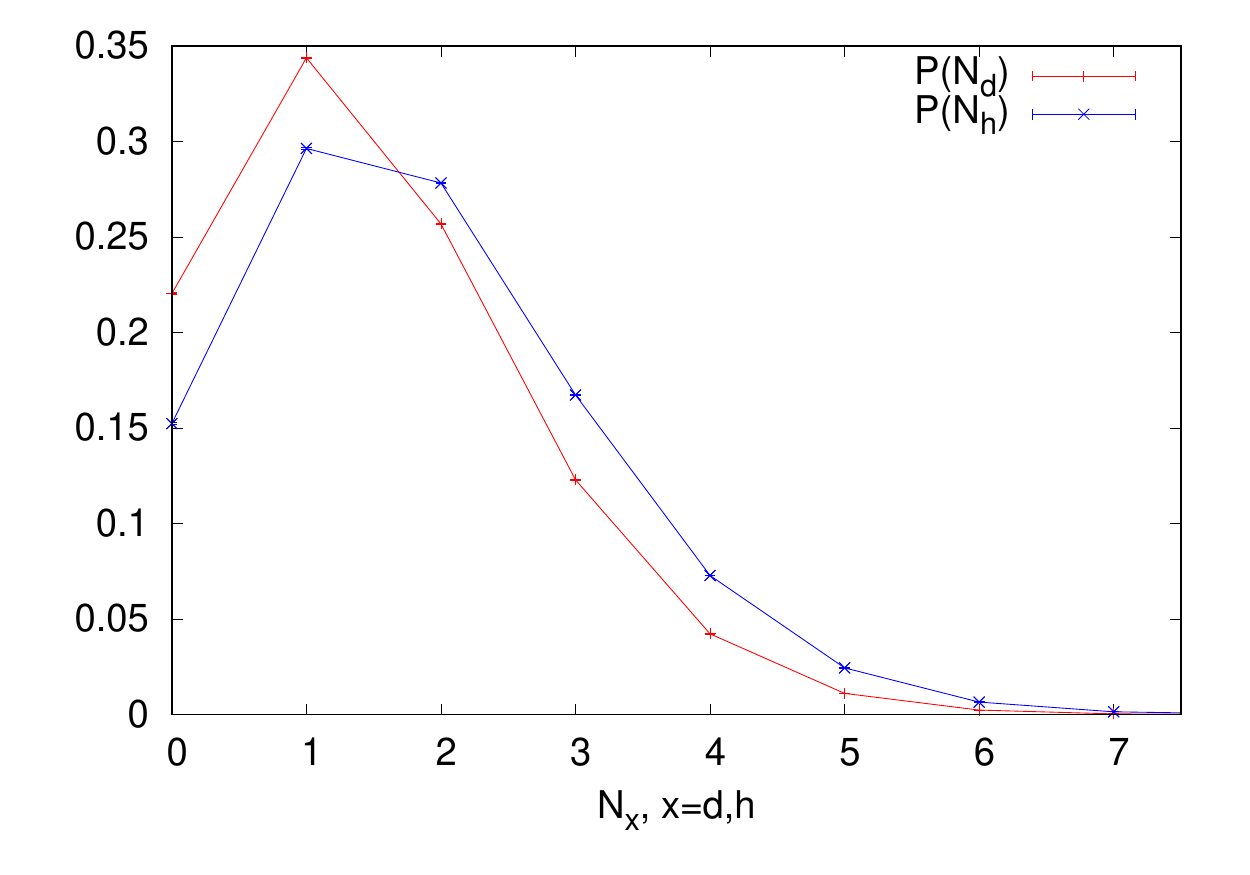}
\caption{FCS of the number of doubly occupied ($N_d$) and empty sites ($N_h$) for the same parameters 
as in Fig.~3 of the main text, which are $L=10, U/t=14, \mu/t = -3.0, \beta t = 2.5$. }
\label{fig:FCS_same_params_as_maintext}
\end{figure}
This section illustrates the significance of numerical simulations for exploring the parameter space. 
In agreement with Ref.~\cite{Blomquist2019AbInitio} for the $t-J$ model and the results of Fig.~3 for the Hubbard model,
one can identify the joint condition $C_{|\vecbf{d}|=2}(|\vecbf{r}|=0) < 0$ and $C_{|\vecbf{d}|=2}(|\vecbf{r}|=1) > 0$ as a characteristic correlation pattern of a magnetic polaron. 
However, $C_{|\vecbf{d}|=2}(|\vecbf{r}|=0) < 0$ by itself occurs also for larger doping values where the antiferromagnetic spin background is too diluted to host polarons. 
Fig.~\ref{fig:no_polaron} juxtaposes the evolution of the three-point correlation function in the reference frame of a hole
\begin{equation}
C_{|\vecbf{d}|=2}(|\vecbf{r}|) = \frac{\langle n_{\vecbf{r}_0}^{(h)} S^{z}_{\vecbf{r}_0 + \vecbf{r} + \frac{\vecbf{d}}{2}} S^{z}_{\vecbf{r}_0 + \vecbf{r} - \frac{\vecbf{d}}{2}}\rangle }{ \langle n_{\vecbf{r}_0}^{(h)} \rangle}. 
\label{eq:cond_spin_charge_no_projector}
\end{equation}
as a function of chemical potential $\mu/t$
for $L=4$ (c) and $L=6$ (e), respectively, with the distribution of the number of holes (b,g) and doubly occupied sites (a,f). 
Unlike in Eq.~(12) of the main text, for the study of the hole environment in Fig.~\ref{fig:no_polaron} 
no projection operator restricting the sites around the hole to be singly occupied has been applied 
prior to evaluating the correlation function.
The data points connected by a thick red line are taken from Fig.~4(c) of Ref.~\cite{Koepsell2019}.
From $\mu/t=-4.0$ to $\mu/t = -5.5$ the correlator $C_{|\vecbf{d}|=2}(|\vecbf{r}|)$ changes markedly
which leads to the conclusion that Fig.~4 of Ref.~\cite{Koepsell2019} is consistent 
with the disappearance rather than the presence of magnetic polarons, which is due to the relatively high level of doping chosen in Ref.~\cite{Koepsell2019}.
This picture is supported by visually comparing the two randomly selected pseudo-snapshots for $\mu/t=-4.0$ versus $\mu/t=-5.0$ in
Fig.~\ref{fig:no_polaron}(d). However, note that the pseudo-snapshots should not be taken at face value
since they come with a sign and a reweighting factor.
Fig.~\ref{fig:FCS_same_params_as_maintext}
shows the FCS of the number of doubly occupied sites and holes for the same parameters 
as in Fig.~\ref{fig:three_point_spincharge_corr} of the main text, indicating that only a small number of excess holes is present
in the pseudo-snapshots.

\end{document}